\newtheorem{theorem}{Theorem}
\newtheorem{proposition}{Proposition}

\documentclass[aip,jmp,reprint,numerical,10pt,floatfix]{revtex4-1}
\usepackage[dvips]{graphics}
\usepackage{graphicx}
\usepackage{amsfonts}
\usepackage{amssymb}
\usepackage{amsmath}
\usepackage{subfigure}
\begin{document}

\title{Stationary states of a nonlinear Schr\"{o}dinger lattice with a harmonic trap}

\author{V. Achilleos}
\affiliation{Department of Physics, University of Athens, Panepistimiopolis, Zografos, Athens 15784, Greece}
\author{G. Theocharis}
\affiliation{Graduate Aerospace Laboratories (GALCIT) California Institute of Technology, Pasadena, CA 91125, USA}
\author{P.G. Kevrekidis}
\affiliation{Department of Mathematics and Statistics, University of Massachusetts, Amherst MA 01003-4515, USA}
\author{N.I. Karachalios}
\affiliation{Department of Mathematics, University of the Aegean, Karlovassi, 83200 Samos, Greece}
\author{F.K. Diakonos}
\affiliation{Department of Physics, University of Athens, Panepistimiopolis, Zografos, Athens 15784, Greece}
\author{D.J. Frantzeskakis}
\affiliation{Department of Physics, University of Athens, Panepistimiopolis, Zografos, Athens 15784, Greece}

\begin{abstract}

We study a discrete 
nonlinear Schr\"{o}dinger lattice
with a
parabolic trapping potential. The model, describing, e.g., an array of repulsive Bose-Einstein condensate droplets confined in the wells of an optical lattice, is
analytically and numerically investigated.
Starting from the linear limit of the problem, we use global bifurcation theory to rigorously prove that -- in the discrete regime -- all linear states lead to nonlinear generalizations thereof, which assume the form of a chain of discrete dark solitons (as the density increases).
The stability of the ensuing nonlinear states is studied and it is found that the ground state is stable, while
the excited states
feature a chain of stability/instability bands.
We illustrate the mechanisms under which discreteness destabilizes the
dark-soliton configurations,
which become stable only inside the continuum regime.
Continuation from the anti-continuum limit is also considered, and a rich bifurcation structure is revealed.

\end{abstract}

\maketitle

\section{Introduction}

The experimental realization of atomic Bose-Einstein condensates (BECs) \cite{FirstBECs} has triggered an intense activity in the study of purely quantum systems at almost macroscopic scales. From a theoretical standpoint,
many effects related to BEC physics can be described by lowest-order mean-field theory, namely the Gross-Pitaevskii equation (GPE) \cite{book2,BECBOOK}. The latter is a nonlinear Schr\"{o}dinger (NLS) equation, incorporating an external trapping potential, with the nonlinearity effectively accounting for inter-atomic interactions. In the absence of the nonlinear term, the GPE becomes a linear Schr{\"o}dinger equation for a confined single-particle state; in this limit, and in the case of, e.g., a harmonic external potential, the linear problem becomes the equation for the quantum harmonic oscillator characterized by discrete energies and corresponding eigenstates \cite{landau}. Then, one may generalize this simple physical picture, and regard the GPE as a model for a {\it self-interacting macroscopic quantum oscillator}; in such a case, the use of analytical and/or numerical techniques for the {\it continuation} of these linear eigenstates (supported by the particular type of the external trapping potential) leads to purely {\it nonlinear states of the self-interacting quantum oscillator}. Such nonlinear states can be found both in one-dimensional (1D) \cite{KivsharPLA,konotop1,zezy1,zezy2} and higher-dimensional settings \cite{kivshar,gregh,mtol,mtolk,zezy3}. Notice that in the 1D setting, and for BECs with repulsive interatomic interactions, the nonlinear states assume the form of dark solitons, which have been studied extensively both in nonlinear optics \cite{kld} and the physics of atomic BECs \cite{djf}.

In this work, we consider and analyze the discrete version of the GPE model,
namely a {\it discrete NLS (DNLS) equation} \cite{panbook}, which incorporates a (discrete) harmonic potential. This model is motivated by the physical setting of a BEC confined in a combined non-negligible harmonic trap and strong periodic potential, the so-called {\it optical lattice}, where rich physical properties and nonlinear dynamics have been revealed \cite{BECBOOK,pandim,OL1,OL2}. Optical lattices are generated by a pair of laser beams forming a standing wave which induces a periodic potential; thus, for a BEC loaded in an optical lattice, the trapping potential in the GPE can be regarded as a superposition of a harmonic trap and a periodic potential. If the harmonic potential is very weak as compared to the optical lattice, it can approximately be ignored; then, the stationary states of the GPE (which includes solely the periodic potential) can be found in the form of nonlinear Bloch waves, which have the periodicity of the optical lattice (see, e.g., Ch.~6 in Ref.~\cite{BECBOOK} and references therein). In the same case (i.e., in the absence of the harmonic potential), if the optical lattice is sufficiently deep (compared to the chemical potential), the strongly spatially localized wavefunctions at the lattice sites can be approximated by Wannier functions (see, e.g., Ref.~\cite{trombettoni1}) and the {\it tight-binding approximation} can be applied; then, the continuous GPE is reduced to the DNLS equation \cite{BECBOOK,pandim,OL1,trombettoni1}, a model which has already a long history in the physics and mathematics of nonlinear lattices \cite{panbook}.
%
Notice that the validity of this model assumes intra-well phase coherence and, thus, it
cannot be used in situations such as the superfluid-to-Mott insulator phase transition \cite{SMIT} or, generally, when strong correlation effects come into play (see, e.g., the review \cite{mlas}). Nevertheless, the model under consideration, apart from being motivated by the physics of BECs loaded in optical lattices -- where it can be regarded as a macroscopic quantum harmonic oscillator on a lattice -- it may also apply in other physical settings, including discrete nonlinear optics \cite{seg} and nonlinear lattice theories \cite{panbook}.

Here, our scope is to study the existence, bifurcations and stability of nonlinear states emerging in this setting for values of the lattice spacing $\alpha$ ranging from the discrete regime ($\alpha ={\rm O}(1)$) to the so-called anti-continuum (AC) limit ($\alpha \rightarrow \infty$). First,
we revisit the linear limit of the problem
(studied some time ago in Ref.~\cite{cgm86}) and generalize the corresponding linear considerations towards finding analytically and numerically the nonlinear states supported by the system. Then, we use global bifurcation theory \cite{RB71,ZeiV1} to rigorously prove that -- in the discrete regime -- each linear eigenstate of the system can be continued to a
nonlinear counterpart. This way, we find all such
nonlinear states, namely the ground state, as well as excited states which, within the strongly nonlinear regime, acquire the form of a chain of dark solitons.
We also study the continuation from the AC limit, through a detailed numerical bifurcation analysis, and find that
there exist states without a linear counterpart.

Furthermore, we study the stability of the nonlinear states in the framework of the linear stability analysis [so-called, Bogoliubov - de Gennes (BdG) analysis, in the realm of BECs]
focusing on the effect of discreteness.
We reveal a fundamental difference of the discrete system
and its continuum counterpart: we find that the discreteness renders the excited states more unstable (only the ground state is found to be always stable)
through the emergence of instability bands; the pertinent band structure depends on the lattice spacing and the nonlinearity strength (as measured by the chemical potential $\mu$).
Contrary, in the continuum case,
not only the ground state but even some of
the excited states (such as the first and second ones)
are
stable deeply inside the nonlinear regime \cite{multi}.
%
We also perform numerical simulations to follow the evolution of the first two (unstable) excited states, namely of the single discrete dark soliton and of the dark soliton pair. We show that the (oscillatory) instability thereof manifests itself by setting a quiescent dark soliton configuration into an oscillatory motion, which can be
explained by resonance effects between the eigenfrequencies of the soliton modes and the intrinsic excitation frequencies of the underlying system.

The paper is organized as follows. In Section II, we present the model as
motivated by the physics of BECs loaded in optical lattices, although, as indicated above, our considerations can be relevant to other fields of applications such as nonlinear optics. In Section III, we also study analytically and numerically the linear limit of the model. In Section IV, we consider the fully nonlinear problem and show, in particular, how nonlinear eigenstates emerge from linear ones; the anti-continuum limit of the system is also studied. In Section V, we analyze the stability of the excited nonlinear states (i.e., the single dark soliton and the two dark soliton states) and, finally, in Section VI, we summarize our conclusions.

\section{Model and methods}

\subsection{Physical motivation and the model}

We consider an atomic BEC confined in a highly anisotropic harmonic trap, $V_{HT}$, with frequencies $\omega_x$ and $\omega_{\perp} \equiv \omega_y=\omega_z$, such that $\omega_x \ll \omega_{\perp}$. In the mean-field approximation, and for sufficiently low temperatures (so that thermal and quantum fluctuations can be neglected), the BEC dynamics can be described by
the following effectively one-dimensional (1D) GPE \cite{BECBOOK},
\begin{eqnarray}
i\hbar\frac{\partial\Psi}{\partial t}=-\frac{\hbar^2}{2m} \frac{\partial^2 \Psi}{\partial x^2}
+ V_{ext}(x)\Psi +g_{\rm 1D}|\Psi|^2\Psi,
\label{gpe}
\end{eqnarray}
where $\Psi(x, t)$ is the macroscopic BEC wavefunction normalized to the number of atoms, namely $\int |\Psi|^2  dx = N$, while $g_{\rm 1D}=2\hbar \omega_{\perp} a $ is the effectively 1D coupling constant, with $m$ being the atomic mass and $a$ the s-wave scattering length, assumed to be positive (i.e., the interatomic interactions are repulsive). Finally, the external potential, $V_{ext}(x)$, in Eq.~(\ref{gpe}) takes the form
\begin{eqnarray}
V_{ext}(x)\equiv V_{HT}(x) = \frac{1}{2}m \omega_x^2 x^2.
\label{Vext}
\end{eqnarray}
Equation (\ref{gpe}) can be expressed in the following dimensionless form:
\begin{equation}
i \frac{\partial \psi}{\partial t} = - \frac{1}{2} \frac{\partial^2 \psi}{\partial x^2}+ V(x)\psi
+ |\psi|^{2}\psi,
\label{dim1dgpe}
\end{equation}
where $|\psi|^2=2a|\Psi|^2$ is the normalized density, while length, time and energy are respectively measured in units
of $2a$, $a_{\perp}$, $\omega_{\perp}^{-1}$ and $\hbar\omega_{\perp}$; the potential $V(x)$ in Eq.~(\ref{dim1dgpe}) is given by:
\begin{equation}
V(x)=\frac{1}{2} \Omega^2 x^2,
\label{htrap}
\end{equation}
where $\Omega\equiv \omega_x/\omega_{\perp}$ is the normalized harmonic trap strength. Notice that apart from the BEC context, Eq.~(\ref{dim1dgpe}) appears also in studies in the nonlinear optics context (see, e.g., Ref.~\cite{nlox2}): there, $\Psi$ is the normalized electric field envelope, $t$ denotes the propagation direction, while the parameter $\Omega$ accounts for the change in the refractive index of the medium in the $x$-direction (transverse to the propagation).

In our analysis below, we consider the discretized version of Eq.~(\ref{dim1dgpe}), namely the following DNLS model:
\begin{eqnarray}
i\dot{\psi}_j=-\frac{1}{2\alpha^2} \Delta_2 \psi_j + \frac{1}{2} \Omega^2 (\alpha j)^2 \psi_j+|\psi_j|^2 \psi_j,
\label{eq1}
\end{eqnarray}
where the overdot denotes time derivative, $\alpha$ is the lattice spacing, and $\Delta_2 \psi_j \equiv \psi_{j+1}-2\psi_j+\psi_{j-1}$ is the discrete Laplacian. In the {\it continuum limit} of $\alpha \rightarrow 0$,  Eq.~(\ref{eq1}) is reduced to the continuum GP model of Eq.~(\ref{dim1dgpe}). Notice that in the absence of the nonlinear term, Eq.~(\ref{eq1}) is the time-dependent problem for a quantum harmonic oscillator (QHO) on a lattice. On the other hand, in the presence of the nonlinear term, the model can be considered as a self-interacting macroscopic QHO on a lattice.

The DNLS of Eq.~(\ref{eq1}), apart from being interesting in its own right, is also relevant to the physics of atomic BECs confined in strong optical lattices. To further elaborate on the above, let us assume that the external potential $V_{ext}$ in Eq.~(\ref{gpe}) incorporates a periodic optical lattice potential, $V_{OL}$, created by two counter-propagating laser beams
of wavelength $\lambda$ \cite{OL1,OL2}; in such a case, $V_{ext}(x)=V_{HT}(x)+V_{OL}(x)$, with the optical lattice potential being given by:
\begin{eqnarray}
V_{OL}(x)=
V_0\sin^2(kx), \label{Vh}
\end{eqnarray}
where $V_0$ and $k=2\pi/\lambda$ denote the strength and wavenumber of the optical lattice, respectively.
%
Then, considering the special case of a sufficiently strong optical lattice, such that $V_0 \gg \mu$ (where $\mu$ is the chemical potential), we may follow the analysis of Ref.~\cite{trombettoni1} and show that Eq.~(\ref{gpe}) can be approximated by a DNLS model for the wavefunctions $\psi_j(t)$ in the different wells (denoted by the index $j$). Particularly, we assume that the effective harmonic trap frequency at each well,
$\tilde{\omega}_x \equiv \sqrt{2V_0k^2/m}$, is such that
$\tilde{\omega}_x \gg \omega_x$, we may employ the tight-binding approximation and decompose the BEC wavefunction $\Psi(x,t)$ as a sum of the wavefunctions $\Phi_j(x-x_j)$ localized around the center of each well, namely,
\begin{eqnarray}
\Psi(x,t)=\sum_j \psi_j(t) \Phi_j(x-x_j),
\label{ansatz1}
\end{eqnarray}
where the wavefunctions $\Phi_j$ are normalized to unity, and the total number of atoms in the condensate
now reads $N=\sum_j N_j= \sum |\psi_j|^2$ (where $N_j$ is the number of atoms at the well $j$). Substituting
Eq.~(\ref{ansatz1}) into Eq.~(\ref{gpe}), multiplying by $\Psi^*$ and integrating over $x$,
yields the following equation for the wavefunctions $\psi_j$ (see Refs.~\cite{trombettoni1,trombettoni}),
\begin{eqnarray}
i\hbar\frac{\partial\psi_j}{\partial t}= &-&K (\psi_{j+1}+\psi_{j-1}) + E_j \psi_j \nonumber \\
&+& \frac{1}{2}m \left(\frac{\lambda}{2}\right)^2 \omega_x^2 j^2 \psi_j +\tilde{g}|\psi_j|^2 \psi_j.
\label{eq11}
\end{eqnarray}
%
To derive the above equation, we have used the (quasi) orthogonality relation $\int dx \Phi_i\Phi_j \approx \delta_{ij}$,
we have kept only terms including spatial integrals of first-neighbor wavefunctions, and
we neglected terms
proportional to $\int dx \Phi^2_j\Phi^2_{j \pm 1}$ and $ \int dx \Phi^3_j\Phi_{j\pm1}$.
%
The constants $K$
and $E_j$ in Eq.~(\ref{eq11}) are given by:
\begin{eqnarray}
K &\approx& \int dx \left[\frac{\hbar^2}{2m} \frac{\partial \Phi_j}{\partial x}
\frac{\partial \Phi_{j\pm 1}}{\partial x}
+\Phi_j V_{ext}(x)\Phi_{j\pm 1}\right], \label{K} \\
E_j &\approx& \int dx \left[\frac{\hbar^2}{2m} \left|\frac{\partial \Phi_j}{\partial x}\right|^2
+|\Phi_j|^2 V_{ext}(x) \right] \label{Ej},
\end{eqnarray}
while $\tilde{g}=g_{\rm 1D} \int dx \Phi_j^4$. Equation (\ref{eq11}) can readily be made dimensionless measuring length, time and energy in units of the lattice spacing $\alpha =\lambda/2$, $\omega_L^{-1}=\hbar/E_L$, and
$E_L=2E_R=\hbar^2/m \alpha^2$ (where $E_R$ is the recoil energy), respectively. In these units, and employing the transformation,
\begin{equation}
\psi_j \rightarrow \sqrt{\frac{\hbar \omega_L}{\tilde{g}}}\psi_j
\exp\left[-i\left(\frac{E_j-2K}{\hbar \omega_L}\right)t\right],
\label{tr}
\end{equation}
we can express Eq.~(\ref{eq11}) as follows:
\begin{eqnarray}
i\dot{\psi}_j=-\epsilon \Delta_2 \psi_j + \frac{1}{2} \tilde{\Omega}^2 j^2 \psi_j+|\psi_j|^2 \psi_j,
\label{eq111}
\end{eqnarray}
where
$\epsilon = K/E_L$ and $\tilde{\Omega}= \omega_x/\omega_L$, respectively.

Formally speaking, the DNLS Eq.~(\ref{eq111}) is a variant of Eq.~(\ref{eq1}), but there are also some differences arising from the dependence of the trap strengths and coefficients of the kinetic terms on the lattice spacing $\alpha$. From a physical viewpoint, Eq.~(\ref{eq111}) applies for the regime corresponding to moderate values of $\alpha$: this is due the fact that the assumptions for the derivation of Eq.~(\ref{eq111}) become invalid for small or large values of the lattice spacing. Nevertheless, it can be found that there exists a certain range of $\alpha$-values (for a given harmonic trap strength $\Omega$), where
Eq.~(\ref{eq111}) is equivalent to Eq.~(\ref{eq1}) -- the formal discretization of Eq.~(\ref{dim1dgpe}): using experimentally relevant parameters \cite{anker} for a rubidium condensate confined in a trap with frequencies $\omega_x = 2\pi \times 10$~Hz and $\omega_{\perp} = 2\pi \times 100$~Hz,
and total number of atoms $N \approx 2000$, one may find that the ratio $\tilde{\Omega}^2/K$ takes values in the interval $0.001 \lesssim \tilde{\Omega}^2/K \lesssim 0.01$. Accordingly, for the fixed value of the normalized trap strength $\Omega=0.1$ (which will be used below), if the lattice spacing takes values in the interval $0.25 \lesssim \alpha \lesssim 0.6$, then Eqs.~(\ref{eq1}) and (\ref{eq111}) become equivalent.

Thus, the model Eq.~(\ref{eq1}) is related to the continuum model Eq.~(\ref{dim1dgpe}) (for $\alpha \rightarrow 0$), describing dynamics of harmonically confined BECs or dynamics of beams in graded-index waveguides, while it can also be used -- in the strongly discrete regime ($\alpha \lesssim 1$) -- to describe the dynamics of arrays of BECs in optical lattices.

It should be noted in passing that, in what follows in our analysis, as the number of atoms tends to zero, quantum effects considered in a number of recent works \cite{carrm1,carrm2,lewen} become important; in such a case, applicability of the mean-field approximation becomes questionable. Nevertheless, our aim here is to utilize the model at hand -- as a relevant mathematical limit -- which can be explored to identify the nonlinear states emerging from the linear ones in the regime where the mean-field description is the appropriate one (i.e., for sufficiently large atom numbers).

\subsection{Stability analysis approach}

Below, we will present results concerning the stability of nonlinear states of Eq.~(\ref{eq1}). In fact, we will perform a linear stability analysis based on the so-called (in the context of BECs) BdG equations \cite{book2,BECBOOK}. In particular,
once a real, stationary state, $\psi_j^{(0)}$, is identified by means of a fixed point algorithm (e.g., a Newton-Raphson
method), we consider small perturbations of this state of the form,
\begin{equation}
\psi_j(t)=[\psi_{j}^{(0)}+(u_j e^{-i\omega t}+\upsilon_j^{\ast} e^{i\omega t})]e^{-i\mu t},
\end{equation}
where the asterisk denotes complex conjugation. Substituting this ansatz into Eq.~(\ref{eq1}), and linearizing with respect to $u_j$ and $\upsilon_j$, we obtain the linear stability (BdG) equations
\begin{eqnarray}
\left[\hat{H}-\mu+2|\psi_{j}^{(0)}|^2 \right]u_j+(\psi_{j}^{(0)})^2 \upsilon_j &=&\omega u_j,
\label{BdG1} \\
\left[\hat{H}-\mu+2|\psi_{j}^{(0)}|^2 \right]\upsilon_j+(\psi_{j}^{(0)\ast})^2 u_j&=&-\omega \upsilon_j,
\label{BdG2}
\end{eqnarray}
where $\hat{H}=-(1/2\alpha^2)\Delta_2+\frac{1}{2}\Omega^2(\alpha j)^2$ is the single-particle operator. Solving these equations one can find the eigenfrequencies $\omega \equiv \omega_{r}+i \omega_{i}$ and the amplitudes $u_j$ and $\upsilon_j$ of the normal modes of the system. Note that due to the Hamiltonian nature of the system, if $\omega$ is an eigenfrequency of the Bogoliubov spectrum, so are $-\omega$, $\omega^{\ast}$ and $-\omega^{\ast}$. A stable (unstable) configuration corresponds to $\omega_i =0$ ($\omega_i \ne 0$).

An important quantity resulting from the BdG analysis is the energy carried by the normal mode with eigenfrequency $\omega$. This is given by the following expression:
\begin{equation}
E=\int{dx(|u|^2-|\upsilon|^2)\omega. }
\label{energy}
\end{equation}
In brief, the energy measure of (\ref{energy}) yields the energy difference between a perturbed state
and an equilibrium (fixed point state), as an explicit calculation (see  equations (5.73)-(5.77) of the Ref.\cite{book2})
clearly illustrates. The sign of this quantity, known as {\it Krein sign} \cite{MacKay}, is a topological property. Importantly, if the normal mode eigenfrequencies with opposite energy (Krein) signs are in resonance then, typically, there appear complex frequencies in the excitation spectrum, i.e., a dynamical instability occurs \cite{MacKay}. In order to further elaborate on such a possibility, we note that modes with complex or imaginary frequencies carry zero energy, while {\it anomalous modes} -- associated with the presence of dark solitons in the configuration -- have negative energy (see, e.g., Sec.~5.6 of Ref.~\cite{book2}). The presence of anomalous modes in the excitation spectrum is a direct signature of an energetic instability or, in other words, is an evidence that the stationary state over which the BdG analysis is applied is not the ground state of the system.

The above analysis scheme will be used in Secs.~IV.B and V below.

\section{The linear problem}
\label{sec2}

Let us start our analysis by considering at first the linear counterpart of Eq.~(\ref{eq1}) resulting from
the substitution $\psi_j \rightarrow \psi_j \exp(-iEt)$ (where $E$ denotes the energy), namely,
%
\begin{eqnarray}
-\frac{1}{2\alpha^2} \Delta_2 \psi_j + \frac{1}{2} \Omega^2 (\alpha j)^2 \psi_j=E \psi_j.
\label{eq2}
\end{eqnarray}
The above equation is the discrete version of the eigenvalue problem describing a QHO on a lattice.
The energy spectrum, as well as the profiles of the pertinent eigenstates, will be used below to construct
(numerically) solutions of the full nonlinear problem. Following the methodology devised in Ref.~\cite{cgm86},
it is possible to solve the discrete QHO eigenvalue problem by considering the following (continuous)
Hamiltonian operator acting on the wavefunction $\tilde{\Psi}(x,t)$:
\begin{eqnarray}
\hat{H}\tilde{\Psi} \equiv -A(e^{i\alpha\hat{p}}+e^{-i\alpha\hat{p}})\tilde{\Psi}
+ \frac{1}{2} \Omega^2x^2\tilde{\Psi} = E'\tilde{\Psi},
\label{gall}
\end{eqnarray}
where $\hat{p}=-i\partial/\partial x$ is the momentum operator, $A$ is the tight-binding constant,
$\alpha$ is a constant, and $E'$ is the energy. Equation~(\ref{gall}) is identical to Eq.~(\ref{eq2})
in a discrete coordinate space: indeed, letting $\tilde{\Psi}(x,t) \rightarrow \psi_j(t)$, $x\rightarrow
\alpha j$ and $V(x)\rightarrow \frac{1}{2}\Omega^2(\alpha j)^2$, the translation operators $\exp(\pm i\alpha\hat{p})$ act
on the wavefunctions as $\exp(\pm i\alpha\hat{p})\psi_j=\psi_{j\pm 1}$
and $\alpha$ corresponds to the lattice spacing.
Then, adding on both sides of Eq.~(\ref{gall}) the term $2A\psi_j$, we find
\begin{equation}
-A(\psi_{j+1}+\psi_{j-1}-2\psi_j)+\frac{1}{2}\Omega^2(\alpha j)^2\psi_j=E\psi_j,
\label{qho}
\end{equation}
where we need to identify $A=1/2\alpha^2$, and $E=E'+2A$. We thus need to  solve the continuous
eigenvalue problem of Eq.~(\ref{gall}). This can be done by expressing it in the momentum representation
(i.e., $\hat{p}\equiv p$ and $\hat{x}\equiv i\partial/\partial p$),
where it can be written as the following Mathieu-type equation \cite{cgm86},
\begin{equation}
\frac{d^2\phi(\upsilon)}{d\upsilon^2}+[b-2q\cos(2\upsilon)]\phi(\upsilon)=0,
\label{mathieu}
\end{equation}
where
\begin{equation}
\upsilon=\frac{\alpha p}{2},\quad q=-\frac{8A}{\Omega^2\alpha^2},\quad b=\frac{8 E'}{\Omega^2\alpha^2}.
\end{equation}
and $\phi(\upsilon)$ is the Fourier transform of $\tilde{\Psi}$.
Equation~(\ref{mathieu}) possesses a well-known energy spectrum and solutions (see, e.g., Ref.~\cite{abram}).
Projecting these solutions on the Hilbert space where $x/\alpha=n$ ($n=1,2,...$) will provide us with the solutions
of Eq.~(\ref{gall}).
The solutions of Eq.~(\ref{mathieu}), namely,
\begin{eqnarray}
\phi^{(even)}_{n}(\upsilon)&=&N_e {\rm ce}_{2n}(\upsilon;q), \qquad n=0,1,2,\cdots \label{sec} \\
\phi^{(odd)}_{n}(\upsilon)&=&N_o {\rm se}_{2n}(\upsilon;q), \qquad n=1,2,3,\cdots \label{cec}
\end{eqnarray}
are called \textit{Mathieu functions} and are periodic, of period $\pi$.
For different values of $q$, these solutions correspond to characteristic values of $b$,
namely $\mathcal{A}_{2n}$ and $\mathcal{B}_{2n}$ for the even and odd eigenfunctions, respectively,
from which we deduce the following energy spectrum:
\begin{eqnarray}
E'^{(even)}_n&=&\frac{1}{8}\Omega^2\alpha^2\mathcal{A}_{2n}(q), \label{energan} \\
E'^{(odd)}_n&=&\frac{1}{8}\Omega^2\alpha^2\mathcal{B}_{2n}(q), \label{energbn}
\end{eqnarray}
for the even and odd eigenfunctions, respectively. The energy spectrum has a simple form, both in the
continuum limit, corresponding to $\alpha\rightarrow 0$,
and the anti-continuum limit, corresponding to $\alpha\rightarrow \infty $;
the respective analytical expressions for $\mathcal{A}_{2n}(q)$ and $\mathcal{B}_{2n}(q)$ can be found in Ref.~\cite{abram}.
As is expected, in the continuum limit, we recover the familiar equidistant QHO energy spectrum with
energies
\begin{equation}
E'_n=\left(n+\frac{1}{2}\right)\Omega, \qquad n=0,1,2,\cdots,
\end{equation}
while in the anti-continuum limit, the energy spectrum becomes parabolic and has the form
\begin{eqnarray}
E'_n=\frac{1}{2}\Omega^2 \alpha^2n^2, \qquad n=0,1,2,\cdots . \label{eneranticont}
\end{eqnarray}
The eigenfunctions in the coordinate space can be obtained upon Fourier transforming the Mathieu functions
of Eqs.~(\ref{sec})-(\ref{cec}). As shown in Ref.~\cite{cgm86}, these eigenfunctions are very similar
to the Hermite polynomials and coincide with the latter in the continuum limit.

The analytical results presented above can directly be compared with numerics.
We first study the ground state energy, for two different oscillator frequencies, spanning all the allowable
range of values of $\alpha$, from the continuum limit ($\alpha \rightarrow 0$) to the anti-continuum one
($\alpha \rightarrow \infty$). The energy given by Eq.~(\ref{energan}) has an approximate analytical form, namely,
\begin{equation}
\mathcal{A}_0=-\frac{1}{2}q^2 + \frac{7}{128}q^4 -\frac{29}{2304}q^6 +...,
\label{a0}
\end{equation}
for sufficiently small values of $q$.

\begin{figure}[tbp]
\includegraphics[scale=0.35]{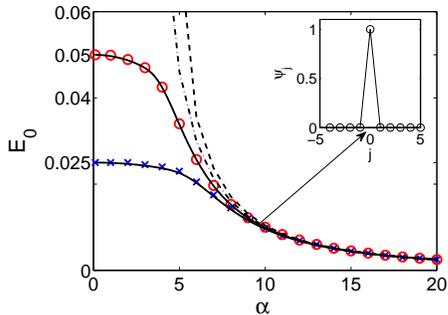}
\caption{(Color online) The ground-state energy $E_0$ for the linear quantum harmonic oscillator as
a function of the lattice spacing $\alpha$, for two values of the trap frequency, $\Omega=0.1$ and
$\Omega=0.05$.
Solid and dotted lines show the energy spectrum as found by solving the QHO eigenvalue problem, while
(red) circles and (blue) crosses show the respective solutions obtained from the Mathieu equation
(\ref{mathieu}). The dashed and dashed-dotted lines show the analytical result of Eq.~(\ref{a0}).
The inset shows the wavefunction profile for $\alpha=10$.}
\label{harmonic}
\end{figure}

In Fig.~\ref{harmonic} we compare the dependence of the ground-state energy on the lattice spacing
$\alpha$ found by numerically solving the QHO eigenvalue problem, with the one obtained by solving the
Mathieu equation (\ref{mathieu}); we also show the approximate analytical result of Eq.~(\ref{energan}).
The analytical solution is only a good approximation for sufficiently large $\alpha$ -- or for small
values of the parameter $q$. As expected, the ground state energy in the continuum limit,
$\alpha \rightarrow 0$, is equal to $\Omega/2$, while in the anti-continuum limit,
$\alpha \rightarrow \infty$, the energy is independent of the trap strength $\Omega$
[cf. Eq.~(\ref{eneranticont})]. The latter result can be understood from the profile of the wave
function in the anti-continuum limit, as seen in Fig.~\ref{harmonic}. In this limit, the only
excited site is $j=0$ which, according to Eq.~(\ref{eq2}), yields $E=\alpha^{-2}$
and asymptotically goes to zero.

\begin{figure}[tbp]
\includegraphics[scale=0.3]{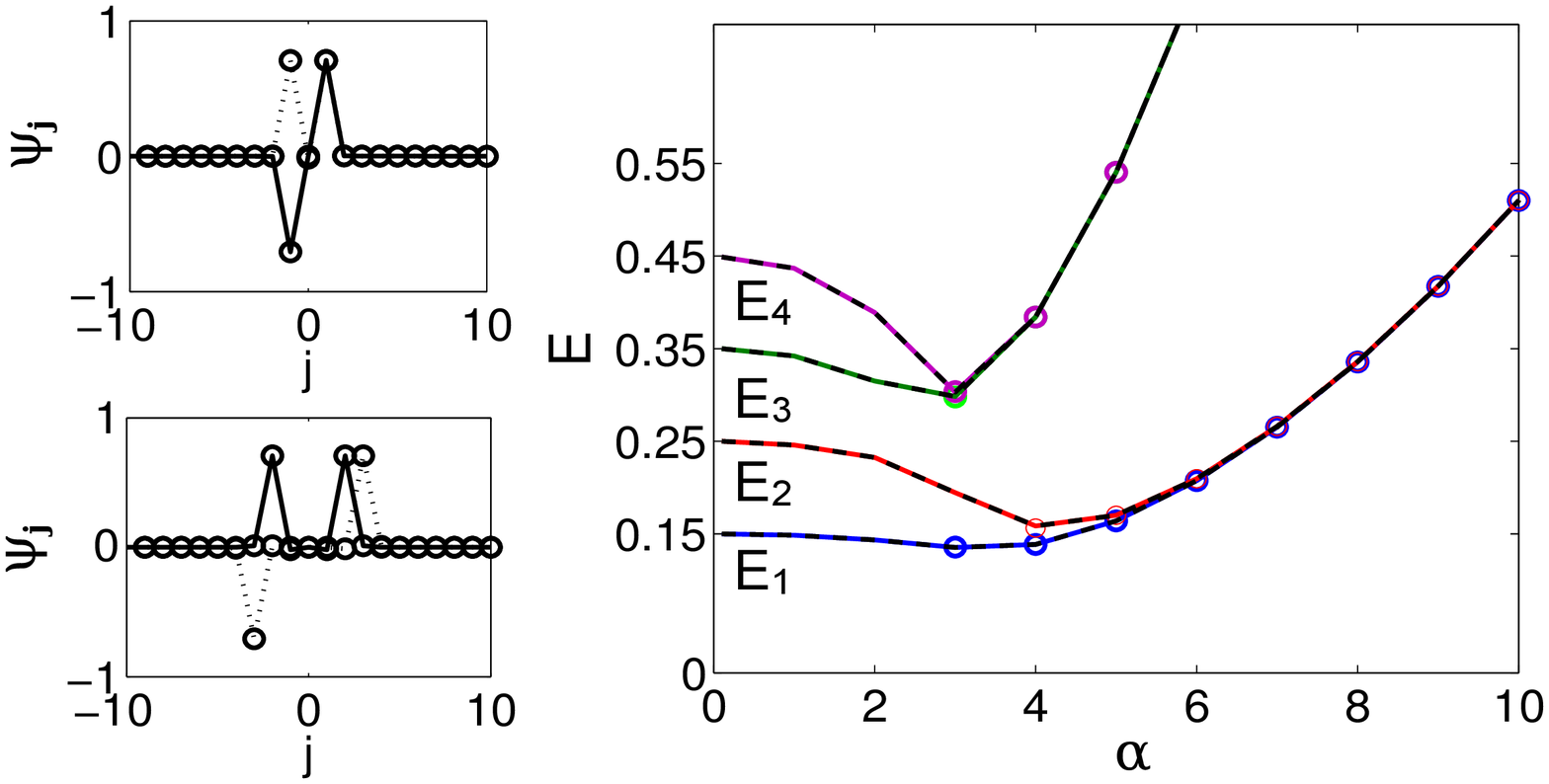} \\
\includegraphics[scale=0.3]{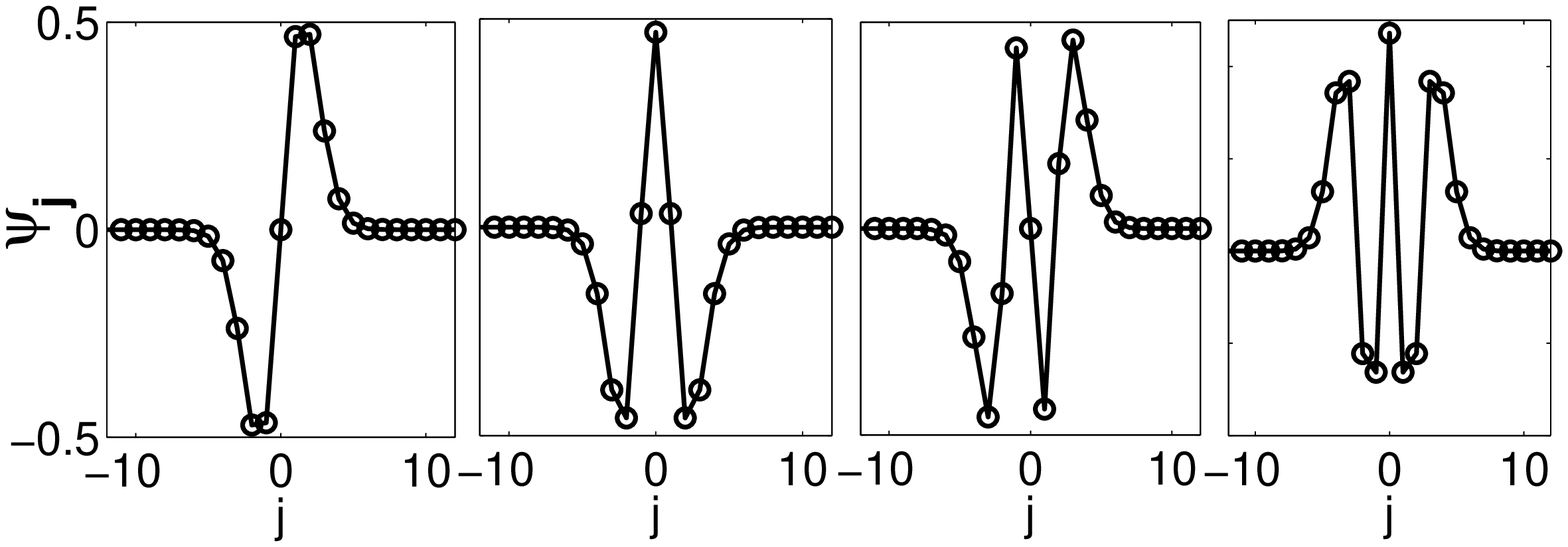}
\caption{(Color online) Top right panel: Energy of the lowest four excited states as functions of the lattice
spacing $\alpha$, for a trap strength $\Omega=0.1$.
Solid lines (blue, red, green, magenta, correspond to the 1st-, 2nd-, 3d-, and 4th-excited states,
respectively) indicate the energy obtained by solving the QHO eigenvalue problem, dashed lines show
the respective solutions obtained from the Mathieu equation (\ref{mathieu}), and circles show the
respective analytical results of Eq.~(\ref{energbn}).
Top left panels: spatial profiles of the 1st-, 2nd-, 3d-, and 4th-excited states for $\alpha=10$ (corresponding to the anti-continuum limit); solid lines depict the 1st- (top) and 3d- (bottom) excited states, while dotted lines depict
the 2nd- (top) and 4th- (bottom) excited states, respectively.
Bottom panels (from left to right): spatial profiles of the 1st-, 2nd-, 3d-, and 4th-excited states for $\alpha=1$ (corresponding to the discrete regime).
}
\label{excited_1}
\end{figure}

Next, we study the spectrum of the first four excited states. The equidistant spectrum in the
continuum limit -- see Fig.~\ref{excited_1} -- becomes parabolic in the fully discrete case.
In the anti-continuum limit it is observed that the excited states become degenerate in pairs.
Again the  profile of the wave functions explains this result: in this limit, $\epsilon \rightarrow 0$,
as seen from Eq.~(\ref{eq2}) the energy depends solely on the potential which is quadratic. As shown in
the inset in the top panel of Fig.~\ref{excited_1} (where the profiles of the first four excited states are shown for $\alpha=10$),
the energy needed to excite symmetrically or anti-symmetrically (with respect to the center) the first neighboring sites is exactly the same due to the quadratic nature of the potential. On the other hand, in the discrete regime, the wavefunction profiles are characterized by the number of nodes, i.e.,
$n$-nodes for the $n$-th excited state; pertinent profiles, for the first four excited states, are shown in the
bottom panels of Fig.~\ref{excited_1} (for $\alpha=1$, and the same trap strength, $\Omega=0.1$).

\section{Existence and bifurcations of solutions in the fully nonlinear problem}

\subsection{Continuation from the linear to the nonlinear regime}

In this section we will study the fully nonlinear case. Our analysis considers an arbitrary number of $K+2$ oscillators equidistantly occupying an interval $[-L, L]$ of length $2L$, with spacing $\alpha=\frac{2L}{K+1}$. Thus, the oscillators are occupying the points $x_j=-L+j\alpha$, $j=0,1,2,\ldots,K+1$ of the interval $[-L,L]$, discretized as
\begin{eqnarray}
\label{disc1}
-L=x_0<x_1<\ldots<x_{K+1}=L.
\end{eqnarray}
We consider the case of {\em real} discrete wavefunctions. For the discrete wavefunctions at each point $x_j$, $j=0,\ldots,K+1$, of (\ref{disc1}),  we use the standard shorthand notation $\psi(x_j)=\psi_j$. In some cases we shall also use the shorthand notation $\psi$ for the vectors of $\mathbb{R}^{K+2}$, i.e., $\psi:=\left\{\psi_j\right\}_{j=0}^{K+1}$.

First, we use the transformation
$\psi_j \rightarrow \psi_j \exp(-i\mu t)$ (where $\mu$ is the chemical potential) to reduce Eq.~(\ref{eq1})
to its time-independent counterpart,
\begin{eqnarray}
-\frac{1}{2\alpha^2}\Delta_2 \psi_j + \frac{1}{2} \Omega^2 (\alpha j)^2 \psi_j +|\psi_j|^2 \psi_j=\mu \psi_j,
\label{eq3}
\end{eqnarray}
for $j=1,\ldots,K$, and assume that the wavefunctions are satisfying Dirichlet boundary conditions at the endpoints $x_0=-L$ and $x_{K+1}=L$, namely:
\begin{eqnarray}
\psi_{0}&=&\psi_{K+1}=0.
\label{dbc}
\end{eqnarray}
Our aim is to use the solutions of the linear problem obtained in the previous section in order to find pertinent solutions in the nonlinear regime. Our analysis starts by proving that the energy spectrum of the nonlinear equation arises from the relevant spectrum found in the linear case. Before proceeding further, it is relevant to note that Eq.~(\ref{eq3}), with the boundary conditions (\ref{dbc}), possesses two conserved quantities: the Hamiltonian $H$ (the energy of the system) and the number of atoms $N$, respectively given by:
\begin{eqnarray}
H&=&\frac{1}{2} \alpha \sum^K_{j=1}\left[\frac{1}{\alpha^2}|\psi_j-\psi_{j-1}|^2
+|\psi_j|^4+ \Omega^2 (\alpha j)^2|\psi_j|^2\right],
\nonumber \\
\label{hamiltonian} \\
N&=&\alpha \sum^K_{j=1} |\psi_j|^2.
\label{Number}
\end{eqnarray}
Notice that the presence of the prefactor $\alpha$ in the definitions of $H$ and $N$ suggests that
in the limit of $\alpha \rightarrow 0$ Eqs.~(\ref{hamiltonian}) and (\ref{Number}) provide the respective Hamiltonian and number of atoms of the continuum GPE, Eq.~(\ref{dim1dgpe}).

The continuation to the nonlinear regime from the linear states (\ref{eq2}), i.e., the bifurcations of
solutions of Eq.~(\ref{eq3}) from solutions of the corresponding linear problem, cf.~Eq.~(\ref{eq2}),
can be justified analytically by using global bifurcation theory -- see Refs.~\cite{RB71} and Section
15.7 of Ref.~\cite{ZeiV1}.
In this setting, we will apply the global bifurcation theorem of Rabinowitz (see Theorem~1.3, p.~490,
of Ref.~\cite{RB71} and Theorem~15.C, p.~668, of Ref.~\cite{ZeiV1}), which we now recall for reasons of completeness
\begin{theorem} \label{rab}
Assume that $X$ is a Banach space with norm\ $||\cdot||_X$.  Consider
the map $\mathcal{F}(\mu,\cdot): X\rightarrow X$, $\mu\in\mathbb{R}$,
$\mathcal{F}(\mu,\cdot)=\mu \mathcal{L}\cdot + \mathcal{H}(\mu,\cdot)$,
where $\mathcal{L}: X\rightarrow X$ is a compact linear map and
$\mathcal{H}(\mu,\cdot): X\rightarrow X$ is compact and satisfies
\begin{eqnarray}
\label{Order}
\lim_{||u||_X \to 0} \frac{||\mathcal{H}(\mu,u)||_{X}}{||u||_X}=0.
\end{eqnarray}
If $\frac{1}{\lambda^*}$ is a simple eigenvalue of $\mathcal{L}$, then the closure of the set
\begin{eqnarray*}
&&C=\{ (\mu ,u) \in \mathbb{R} \times X : (\mu,u)\;\;\\
&&\mbox{solves}\;\; u-\mathcal{F}(\mu,u)=0,\; u \not\equiv 0 \},
\end{eqnarray*}
possesses a maximal continuum (i.e. connected branch) of
solutions $C$ which branches out of $(\lambda^*,0)$ and $C$ either:

(i)\ meets infinity in\ $\mathbb{R} \times X$\ or,

(ii)\ meets\ $u=0$ in a point $(\hat{\mu},0)$ where $\hat{\mu}\neq\lambda^*$ and $\frac{1}{\hat{\mu}}$
is an eigenvalue of $\mathcal{L}$.
\end{theorem}
To apply Theorem \ref{rab}, we need some preparations, in order to rewrite (\ref{eq3}) in the form
$\psi-\mu \mathcal{L}(\psi) + \mathcal{H}(\mu,\psi)=0$ requested by the theorem. As a first step, we will define and discuss the properties of the linear operator $\mathcal{L}$ through the linear eigenvalue problem (\ref{eq2})-(\ref{dbc}), which is the eigenvalue problem for the linear operator
\begin{eqnarray}
\label{bif1}
\mathcal{T}(\psi)_j=-\frac{1}{2\alpha^2}\Delta_2\psi_j+\frac{1}{2}\Omega^2 (\alpha j)^2\psi_j,
\end{eqnarray}
$j=1,\ldots,K$, supplemented with the Dirichlet boundary conditions (\ref{dbc}).
We shall also discuss some properties of the eigensolutions of (\ref{bif1}) related to the number of sign-changes of the discrete eigenfunctions. These properties will be useful for distinguishing between the possibilities (i) and (ii) described by Theorem \ref{rab}.
As a second step, we will define the nonlinear operator $\mathcal{H}$ through the nonlinearity of (\ref{eq3}).

The operator (\ref{bif1}) is strongly positive and selfadjoint on the Hilbert space
$$X=\{\psi=\{\psi_j\}_{j=0}^{j=K+1}\in \mathbb{R}^{K+2}\;:\;\psi_{0}=\psi_{K+1}=0\},$$
having the role of the Banach space $X$  which is referred in Theorem \ref{rab}. The Hilbert space $X$ is endowed with the norm (\ref{Number}), i.e.,
\begin{eqnarray*}
||\psi||_X^2=&\alpha \sum^{K+1}_{j=0} |\psi_j|^2=\alpha \sum^{K}_{j=1} |\psi_j|^2=N.
\end{eqnarray*}
The operator $K$ possesses $K$ simple eigenvalues,
\begin{eqnarray}
\label{bif4}
0<E_0<E_2<\ldots<E_{K-1}.
\end{eqnarray}
Furthermore, the Krein-Rutman theorem (see p.~122 of Ref.~\cite{Smo94}, and Section~7.8, p.~289 of
Ref.~\cite{ZeiV1}), implies that the principal eigenstate associated to the principal eigenvalue
$E_0$ is positive in the sense that $\psi^0_j\geq 0$ for all $j=0,\ldots,K+1$, $\psi^0$ has at least one
positive coordinate and satisfies the boundary conditions (\ref{dbc}).
On the other hand, the eigenvalue problem for the operator (\ref{bif1}) with the boundary conditions (\ref{dbc}), is the discrete analogue of the
Sturm-Liouville problem for the QHO
\begin{eqnarray}
\label{SLGP1}
-\frac{1}{2}\psi''(x)+\frac{1}{2}\Omega^2 x^2 \psi(x)&=&\lambda \psi,\;\;-L<x<L\\
\label{SLGP2}
\psi(-L)&=&\psi(L)=0,
\end{eqnarray}
for which the classical Sturm-Liouville theorem holds (see p.~454 of Ref.~\cite{CH53}). For instance,
(\ref{SLGP1})-(\ref{SLGP2}) has a countable sequence of eigenvalues $\lambda_0<\lambda_2<\ldots ,$ with
corresponding eigenfunctions $\psi_0(x),\psi_1(x),\ldots ,$ and the eigenfunction $\psi_n(x)$, $n=0,1,\ldots ,$ has exactly
$n$ zeros (nodal points) on $(-L, L)$. Continuing the discussion from the end of Section \ref{sec2}, the discrete eigenstates $\psi^n$, $n=0,1,\ldots , K-1$ corresponding to the eigenvalues (\ref{bif4})
interpolate the  continuous eigenfunctions
$u_0(x)$, $n=0,1,\ldots, K$, and they have exactly $n$ nodal points,
$n=0,1,\ldots ,K-1$. We remark that for any $\epsilon>0$ the interpolation and the ``nodal properties'' of the discrete eigenfunctions of (\ref{bif1}) have been described in Refs.~\cite{cgm86,gc91}. Under this observation, for each $n=0,\ldots,K-1$, we may define the following sets in $X$,
\begin{eqnarray}
\label{sceig}
&&S_{n}:=\{\psi=\{\psi_j\}_{j=0}^{j=K+1}\in \mathbb{R}^{K+2}\;:\;\psi_{0}=\psi_{K+1}=0,\nonumber\\
&&\mbox{and has exactly $n$ nodal points}\}.
\end{eqnarray}
The sets $S_{n}$ are clearly open in $X$, since for any $\psi\in S_{n}$ we may construct an $r$-neighborhood $B(\psi,r):=\left\{\phi\in X\;:\;||\psi-\phi||_X<r\right\}$, lying in $S_n$, by considering $r$ sufficiently small. For instance, for $r$-sufficiently small we get sufficiently small perturbations of the coordinates of $\psi$ in $S_n$ and thus all the vectors of $X$ being in $B(\psi,r)$, have the same number of nodal points (i.e., a small perturbation of $\psi\in S_{n}$ lies in $S_n$).

To conclude with our preparations, we write the nonlinear eigenvalue problem (\ref{eq3}) in the form of an
operator equation in $X$, as follows:
\begin{eqnarray}
\label{bif10a}
\mathcal{T}(\psi)-\mu\psi +\mathcal{F}(\psi)=0,\;\;\psi\in X,
\end{eqnarray}
where $\mathcal{F}:X\rightarrow X$ is the nonlinear operator
\begin{eqnarray*}
\mathcal{F}(\psi)_j=|\psi_j|^2\psi_j.
\end{eqnarray*}
The linear operator $\mathcal{T}:X\rightarrow X$ is invertible. Its inverse $\mathcal{T}^{-1}:X\rightarrow X$
is also symmetric and it readily follows that $\nu_n:=\frac{1}{E_n}$, $n=0,1,\ldots,K-1$, are also simple
eigenvalues of $\mathcal{T}^{-1}:X\rightarrow X$. We may write Eq.~(\ref{bif10a}) as
\begin{eqnarray}
\label{bif10b}
\psi -\mu\mathcal{T}^{-1}(\psi)+\mathcal{T}^{-1}\mathcal{F}(\psi)=0
\end{eqnarray}
Equation (\ref{bif10b}) is actually in the form requested by Theorem \ref{rab}
\begin{eqnarray}
\label{bif10c}
\psi-\mu\mathcal{L}(\psi)+\mathcal{H}(\psi)=0,
\end{eqnarray}
with the linear operator $\mathcal{L}:=\mathcal{T}^{-1}:X\rightarrow X$ and the nonlinear operator
$\mathcal{H}:=\mathcal{T}^{-1}\mathcal{F}:X\rightarrow X$ being compact since they are acting on the
finite dimensional space $X$. The map $\mathcal{F}$ is defined by the cubic nonlinearity and, thus,
it is not difficult to verify that $\mathcal{H}$ satisfies condition (\ref{Order}) of Theorem \ref{rab}.
Hence, all the assumptions of Theorem \ref{rab} are satisfied, {\em justifying that $(E_n, 0)$, $n=0,1,\ldots,K-1$ is a
bifurcation point for the problem (\ref{eq3})}.
We may summarize in the following:
\begin{proposition}
\label{prop1}
For any $\alpha>0$, there exists a maximal continuum (i.e. connected branch) of
solutions $C_{E_n}$ of Eq.~(\ref{eq3}), $n=0,1,\ldots,K-1$, bifurcating from $(E_n,0)$ and $C_{E_n}$ either
(i) meets infinity in $\mathbb{R} \times X$, or
(ii)\ meets $\psi=0$ in a point $(\hat{\mu},0)$ where $\hat{\mu}\neq E_n$ and $\frac{1}{\hat{\mu}}$
is an eigenvalue of $\mathcal{L}$.
\end{proposition}

We proceed by discussing some geometric properties of the branches $C_{E_n}$. Considering the eigenstates
$\psi^n$ of Eq.~(\ref{eq2}) corresponding to the eigenvalues $E_n$, the local bifurcation theory and the implicit function theorem [see \cite[Theorem 13.4 pg. 171 \& Theorem 13.5, pg. 173]{Smo94}] guarantees
that the local branch $C_{E_n}$ can be locally represented by the $C^1$ curve
$$(\mu,\psi):(-\delta,\delta)\rightarrow\mathbb{R}\times X,$$ for some $\delta$ sufficiently small,  such that
\begin{eqnarray}
\label{bif13}
&&\mu(0)=E_n,\;\;\chi(0)=0,\;\; \nonumber \\
&&(\mu(s),\psi(s))=(\mu(s),s(\psi^n+\chi(s))),\;\;|s|<\delta,
\end{eqnarray}
where $||\chi(s)||_X=O(|s|)$, in the neighborhood of the bifurcation point $(E_n, 0)$. Furthermore,
there is a neighborhood of $(E_n, 0)$, such that any zero of the equation (\ref{bif10c}) lies on this curve, or is of the form $(E_n, 0)$).

\begin{proposition}
\label{prop2}
Consider the local representation (\ref{bif13}) of the branch $C_{E_n}$. Then, $\mu'(0)=0$, $\mu''(0)>0$ and
the branch is locally concave up.
\end{proposition}
\textbf{Proof:} We insert the expression $((\mu(s),\psi(s))=(\mu(s), s\psi^n+s\chi(s))$ in Eq.~(\ref{eq3})
and we divide by $s$. Then we obtain the equation (recalling that $\chi_0(s)=\chi_{K+1}(s)=0$),
\begin{eqnarray}
\label{bif14}
\mu(s)(\psi_j^n+\chi_j(s))&=& -\frac{1}{2\alpha^2}\Delta_2(\psi_j^n+\chi_j(s)) \nonumber \\
&&+\frac{1}{2}\Omega^2 (\alpha j)^2(\psi_j^n+\chi_j(s))
\nonumber\\
&&+ s^2|\psi_j^n+\chi_j(s)|^2(\psi_j^n+\chi_j(s)). \nonumber \\
\end{eqnarray}
We now differentiate Eq.~(\ref{bif14}) with respect to $s$, namely,
\begin{eqnarray}
\label{bif15}
&&\mu'(s)(\psi_j^n+\chi_j(s))+\mu(s)\chi_j'(s) =-\frac{1}{2\alpha^2}\Delta_2\chi'_j(s) \nonumber \\
&&+\frac{1}{2}\Omega^2 (\alpha j)^2 \chi'_j(s)
+2s|\psi_j^n+\chi_j(s)|^2(\psi_j^n+\chi_j(s))\nonumber\\
&&+3s^2|\psi_j^n+\chi_j(s)|^2\chi'_j(s),
\end{eqnarray}
and by setting $s=0$ in Eq.~(\ref{bif15}) and using Eq.~(\ref{bif13}), we have:
\begin{eqnarray}
\label{bif16}
-\frac{1}{2\alpha^2}\Delta_2\chi_j'(0)&+&\frac{1}{2}\Omega^2 (\alpha j)^2 \chi'_j(0)\nonumber\\
&&=\mu'(0)\psi_j^n+E_n\chi'_j(0).
\end{eqnarray}
Multiplication of Eq.~(\ref{bif16}) by $\psi^n$ and summation by parts, yields
\begin{eqnarray}
\label{bif17}
&-&\frac{1}{2\alpha^2}\sum_{j=0}^{K+1}\Delta_2\chi'_j(0)\psi_j^n+\frac{1}{2}\alpha^2\Omega^2\sum_{j=0}^{K+1}j^2\chi'_j(0)\psi_j^n
\nonumber \\
&=&-\frac{1}{2\alpha^2}\sum_{j=0}^{K+1}\chi'_j(0)\Delta_2\psi_j^n+\frac{1}{2}\alpha^2\Omega^2\sum_{j=0}^{K+1}\chi'_j(0)j^2\psi_j^n
\nonumber \\
&=&\sum_{j=0}^{K+1}\mu'(0)|\psi^n_j|^2+\sum_{j=0}^{K+1}\chi'(0)_jE_n\psi_j^n
\end{eqnarray}
Since $E_n$ and $\psi_j^n$ solve Eq.~(\ref{eq2}), we have that
\begin{eqnarray*}
&-&\frac{1}{2\alpha^2}\sum_{j=0}^{K+1}\chi'_j(0)\Delta_2\psi_j^n+\frac{1}{2}\alpha^2\Omega^2\sum_{j=0}^{K+1}\chi'_j(0)j^2\psi_j^n \\
&=&\sum_{j=0}^{K+1}\chi'(0)_jE_n\psi_j^n.
\end{eqnarray*}
Thus, from Eq.~(\ref{bif17}) we get that
\begin{eqnarray*}
\sum_{j=0}^{K+1}\mu'(0)|\psi^n_j|^2=0,
\end{eqnarray*}
implying that $\mu'(0)=0$. Next, by differentiating Eq.~(\ref{bif15}) with respect to $s$, and setting $s=0$,
one obtains the equation
\begin{eqnarray*}
&&-\frac{1}{2\alpha^2}\Delta\chi''_j(0)+\frac{1}{2}\Omega^2 (\alpha j)^2 \chi''_j(0)+2|\psi_j^n|^2\psi_j^n \\
&&=\mu''(0)\psi_j^n+E_n\chi''_j(0).
\end{eqnarray*}
Working as before, this time we derive that
$$2\sum_{j=0}^{K+1}|\psi_j^n|^4=\mu''(0)\sum_{j=0}^{K+1}|\psi_j^n|^2.$$
Therefore, $\mu''(0)>0$. $\diamond$

Proposition \ref{prop2}, actually states that at least locally, the graph of the $C^1$ function  $\mu(N)$ ($C^1$ curve) is concave up and monotone, thus locally invertible. By interchanging the axes and plotting the graph  of  $N$ as a function of $\mu$, we recover that $N(\mu)$  has locally the same concavity properties (see, e.g., Ch.~13 of Ref.~\cite{Smo94} for bifurcation diagrams), as stated in Proposition \ref{prop2}.

It remains to show that the branches $C_{E_n}$ are global, i.e., that {\it the option (ii) of
Theorem \ref{rab} should be excluded}.
\begin{theorem}
\label{thg3}
For any $\alpha>0$, the maximal continuum (connected branch) of solutions $C_{E_n}$ of Eq.~(\ref{eq3}) bifurcating from
$(E_n,0)$ meets infinity in $\mathbb{R}$. It is locally concave up and is not possessing a maximum (minimum) point.
\end{theorem}
\textbf{Proof:} (a) Recall that any solution $(\mu, \psi)$ close to $(E_n,0)$ has the same number
of nodal points as the eigenstate $\psi^n$ corresponding to the eigenvalue $E_n$. This is due to the
$C^1$-representation of the solution $\psi$ as $\psi_j(s)=s\psi^n_j+s\chi_j(s)$.
For instance,  each linear state $\psi^n$ belongs to the set $S_n$ defined in Eq.~(\ref{sceig}) and $||\chi(s)||_X=O(|s|)$. It follows
then, that the solution $\psi$ satisfies the estimate
\begin{eqnarray*}
||\psi(s)||_X\leq |s|\,||\psi^n||_X+O(s^2),
\end{eqnarray*}
in the neighborhood of the bifurcation point $(E_n,0)$.  Therefore, since the set $S_n$ is open, we get from the above estimate, that $\psi\in S_n$ for $|s|<\delta$.
(b) Now for all $(\mu,\psi)\in C_{E_n}$ and each $n=0,1,\ldots,K-1$, we consider the indicator function
$$
f(\mu,\psi)=\left\{
\begin{array}{rlr}
&1,\;\;\mbox{if}\;\;\psi\in S_n,\\
&0,\;\;\mbox{if}\;\;\psi=0,\;\;\mu=E_m,\;\;m\neq n.
\end{array}
\right.
$$
that is, $f(\mu,\psi)=0$ if the branch $C_{E_n}$ meets the axis $(\mu,0)$ in another eigenvalue $E_m\neq E_n$.
Note that $f$ is well defined due to the two possibilities described by Theorem \ref{rab}.
From (a) we have that if $(\mu,\psi)$ is in some small neighborhood of $(E_n,0)$, then $f(\mu,\psi)=1$.
Thus, the function $f$ is constant (and equals to $1$) in a small neighborhood of $(E_n,0)$, and cannot change
value in this small neighborhood, i.e., $f$ is locally constant. The set $S_n$ is open and the function $f$ is locally constant on the
connected set $C_{E_n}$. Both facts clearly imply that $f$ is continuous. Therefore, $f(C_{E_n})$ should be
also connected, since the image of a connected set through a continuous function should be connected.
However, $f$ is integer valued, and the fact that $f(C_{E_n})$ is connected, implies that $f$ should be
constant, $f=1$, for all $(\mu,\psi)\in C_{E_n}$. Therefore, $C_{E_n}$ cannot contain a point $(E_m,0)$
with $E_m\neq E_n$ and $C_{E_n}$ should be unbounded.

\begin{figure}[tbp]
\includegraphics[scale=0.35]{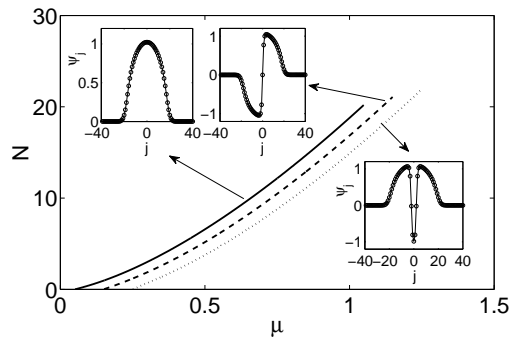}
\caption{The number of atoms $N$ as a function of the chemical potential $\mu$
(for $\alpha=0.8$ and $\Omega=0.1$) for the three lowest states: the ground state (solid line),
the first excited state (dashed line), and the second excited state (dotted line).
Each branch begins from the linear limit ($N=0$), where $\mu$ equals the energy of
the corresponding linear state. The insets show the profiles of these nonlinear states for $\mu=1.2$. }
\label{energyatomsdis}
\end{figure}

Concerning the concavity of the branch, due to Proposition \ref{prop2}, each branch $C_{E_n}$ is concave up
at least for $|s|<\delta$. To prove that is not possessing maximum or minimum points, we will apply
a contradiction argument. Let us assume that the
branch $C_{E_n}$ has a local maximum at some point. Then, due to the $C^1$-property of the branch $C_{E_n}$, and since the branch is connected and unbounded, it follows that $C_{E_n}$ should possess a local minimum. However, as it is shown in Theorem \ref{thresh} in the Appendix,
such a minimum (here possibly attained at some $\mu$), can exist in the case of a DNLS Eq.~(\ref{eq1}) considered in the higher-dimensional lattice $\mathbb{Z}^\mathcal{N}$, $\mathcal{N}\geq 1$,
with power nonlinearity, namely $F(z)=|z|^{2\sigma}z$, only in the case $\sigma\geq \frac{2}{\mathcal{N}}$.
Hence such a minimum in the case of a
$1\mathrm{D}$-lattice can only exist when $\sigma\geq 2$, which is excluded for the time-independent DNLS Eq.~(\ref{eq3})
with the cubic nonlinearity of $\sigma=1$. $\diamond$

We have rigorously proved that a nonlinear state of Eq.~(\ref{gall}) can be created by a continuation
of its linear state in $\mu$. Our analytical results can directly be compared to numerical ones: indeed, using
a Newton-Raphson method, we can construct such nonlinear states starting from their linear counterpart.
In Fig.~\ref{energyatomsdis}, we plot the number of atoms $N=\sum_j|\psi_j|^2$
of the first three states, namely the ground state (solid line), first-excited state (dashed line)
and second-excited state (dotted line), as a function of the chemical potential $\mu$. The corresponding
branches begin from the linear limit ($N=0$), where $\mu$ equals the energy of
the pertinent linear state, and are concave up, in accordance to the analysis presented above.
The insets of Fig.~\ref{energyatomsdis} show the profiles of these nonlinear states for $\mu=1.2$.
It is important to notice that, similarly to the continuous case \cite{konotop1,KivsharPLA,zezyulin,multi},
the excited nonlinear states transform into a chain of discrete dark solitons: the first-excited state
corresponds to a single dark soliton (one node in the wavefunction profile -- see the middle inset of
Fig.~\ref{energyatomsdis}), the second-excited state corresponds to a pair of dark solitons
(two nodes in the wavefunction profile -- see the right inset of Fig.~\ref{energyatomsdis}), and so on.

\subsection{Continuation from the anti-continuum limit}

\begin{figure}[tbp]
\includegraphics[scale=0.32]{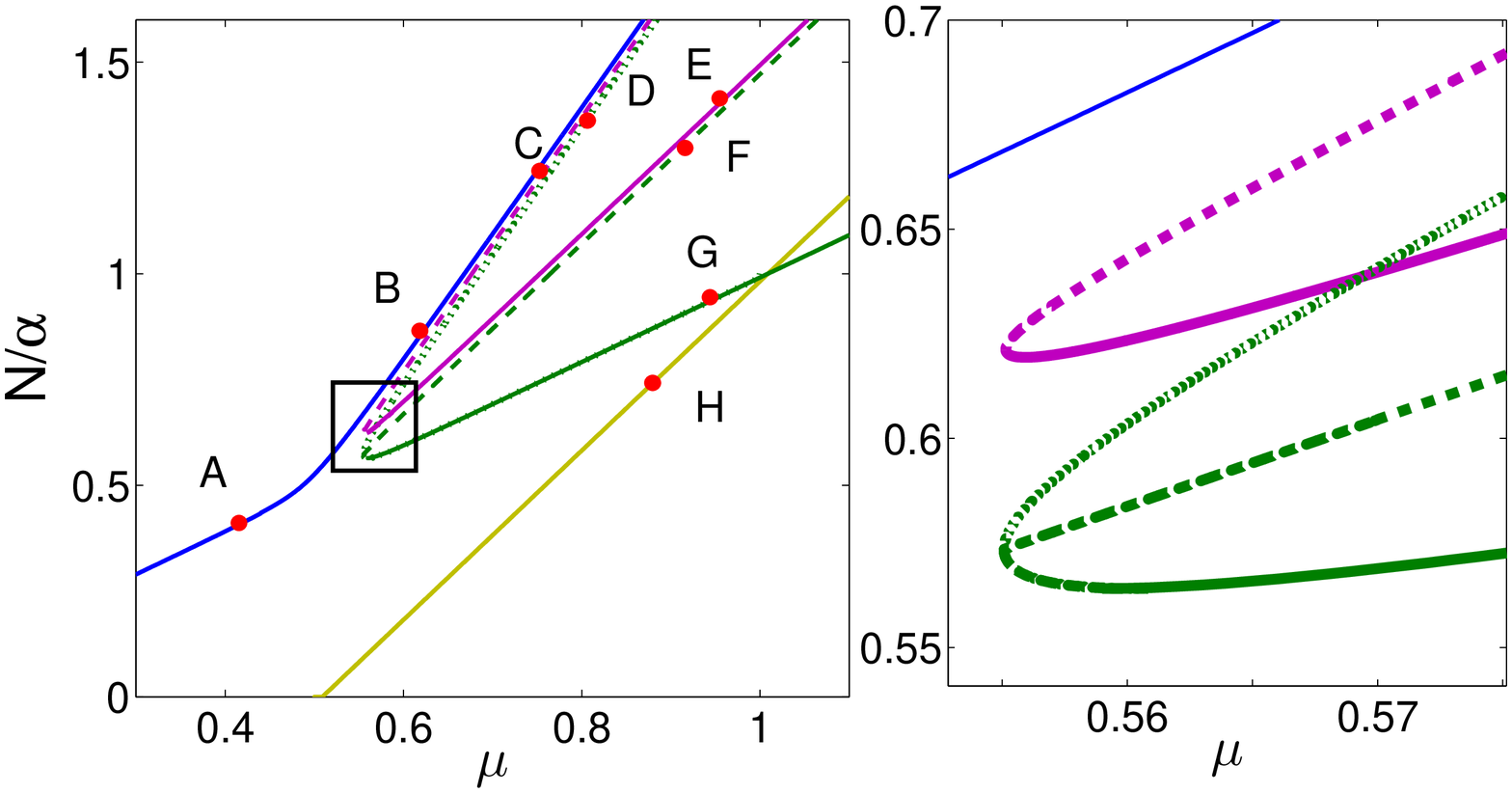}
\includegraphics[scale=0.29]{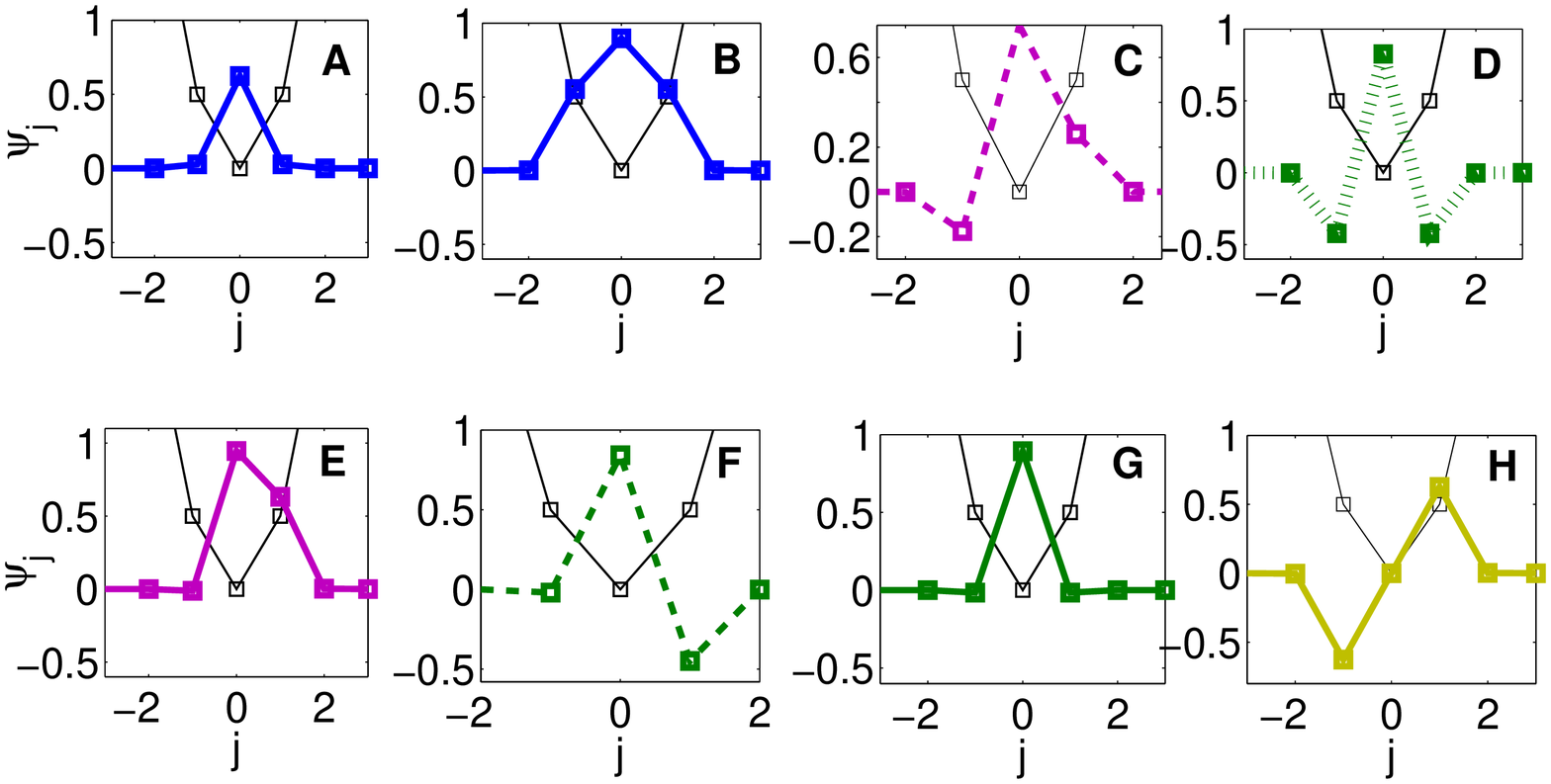}
\caption{(Color online) Top left panel: The normalized number of atoms $N/\alpha$ as a function of the chemical potential $\mu$, for $\alpha=10$ (i.e., in the vicinity of the anti-continuum limit) and $\Omega=0.1$. The black square indicates the region where this panel is magnified, as shown in top right panel. The letters A,B,...,H denote certain points in the diagram for which corresponding wavefunction profiles are shown in the middle and bottom panels. Stable (unstable) branches and respective states are depicted by solid (dashed or dotted) lines.}
\label{energyvsatoms}
\end{figure}

Before discussing in detail the stability of nonlinear states in the form of discrete dark solitons,
in this subsection we will consider the existence and stability of nonlinear states near the AC limit,
in order to appreciate the emerging bifurcation structure.

Near the anti-continuum limit, corresponding to lattice spacing $\alpha \rightarrow \infty$, it is straightforward to find solutions of Eq.~(\ref{eq3}) in the following form:
\begin{equation}
\psi_j = \exp(i\theta_j)\sqrt{\mu-\frac{1}{2}\Omega^2 (\alpha j)^2},
\label{tfdiscrete}
\end{equation}
where $\theta_j$ denotes the phase. The density $|\psi_j|^2$ of the above solution resembles the density profile that can be obtained, in the Thomas-Fermi limit \cite{book2}, from the continuum GPE, Eq.~(\ref{dim1dgpe}).
Thus, all the solutions for any number $n$ of excited sites can
be constructed following Eq.~(\ref{tfdiscrete}).
One can find the analytical expression for the chemical potential with respect to the number of atoms for any such configuration of $n$ excited sites: indeed, using Eq.~(\ref{Number}) and the solutions (\ref{tfdiscrete}), the normalized number of atoms $N/\alpha$ reads:
\begin{eqnarray}
N/\alpha =n \mu-\sum_{j}\Omega^2 (\alpha j)^2,
\label{numbdiscrete}
\end{eqnarray}
where the sum runs over the excited sites. From the above result, it can easily be found that
the slope $\eta \equiv \partial (N/\alpha)/\partial \mu =n$ (for fixed $n$)
does not depend on $j$ -- i.e., which particular sites are excited -- but only on the number of excited sites. In the top panel of Fig.~\ref{energyvsatoms}, we show the dependence of $N/\alpha$ on the chemical potential $\mu$, for states
consisting of up to three excited sites. Note that in this figure we have used the value $\alpha=10$, but we have checked that qualitatively similar results can be obtained for larger values of the  lattice spacing.
As it is observed in the figure, $N/\alpha$ depends linearly on $\mu$ near
the AC limit, in agreement with the analytical prediction of Eq.~(\ref{numbdiscrete}). Notice that the latter is, strictly speaking, valid in the limit of $\alpha \rightarrow \infty$, but the linear dependence of $N/\alpha$ on $\mu$ persists for the chosen finite value of $\alpha$, except at particular slope-changing
critical points explained below.

Let us now describe the result of Fig.~\ref{energyvsatoms} in more detail. We start with the (blue solid line) branch, corresponding to the simplest possible configuration, with only the center site (at $j=0$) excited; an example of a state of this branch is shown in the first of middle panels of Fig.~\ref{energyvsatoms} (state A, with ``A'' in the top left panel marking the respective values of $N/\alpha$ and $\mu$; a similar notation is used for the other branches below). This branch starts from the origin, with slope $\eta =1$, but for values of chemical potential $\mu> \mu_{c}^{(1)} \equiv \frac{1}{2}\Omega^2 (\alpha j)^2 =0.5$ (for $j= \pm 1$) it changes slope, namely $\eta=3$, as two more sites are excited; an example of such a state belonging in this branch for $\mu>\mu_{c1}$ is state B (second middle panel of Fig.~\ref{energyvsatoms}). Notice that further increase of $\mu$ results in a similar behavior for higher values of $j$ (not shown), i.e., this branch changes slope at characteristic values of the chemical potential $\mu_{c}^{(m)} \equiv \frac{1}{2}\Omega^2 (\alpha j)^2$ (for $j= \pm m$), as more sites are excited. This (ground state)
branch of solutions is found to be stable throughout its continuation.
\begin{figure}[tbp]
\includegraphics[scale=0.4]{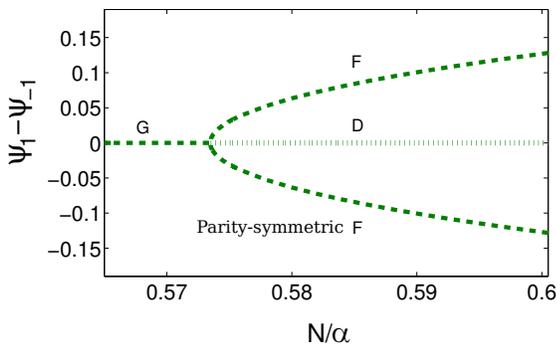}
\caption{(Color online) The pitchfork bifurcation, relevant to the (green) branches D, F (and its parity-symmetric one) and G, as viewed by the difference $\psi_1-\psi_{-1}$ of the two outer sites as a function of the normalized atom number $N/\alpha$.}
\label{beef}
\end{figure}

\begin{figure*}[tbp]
\includegraphics[scale=0.32]{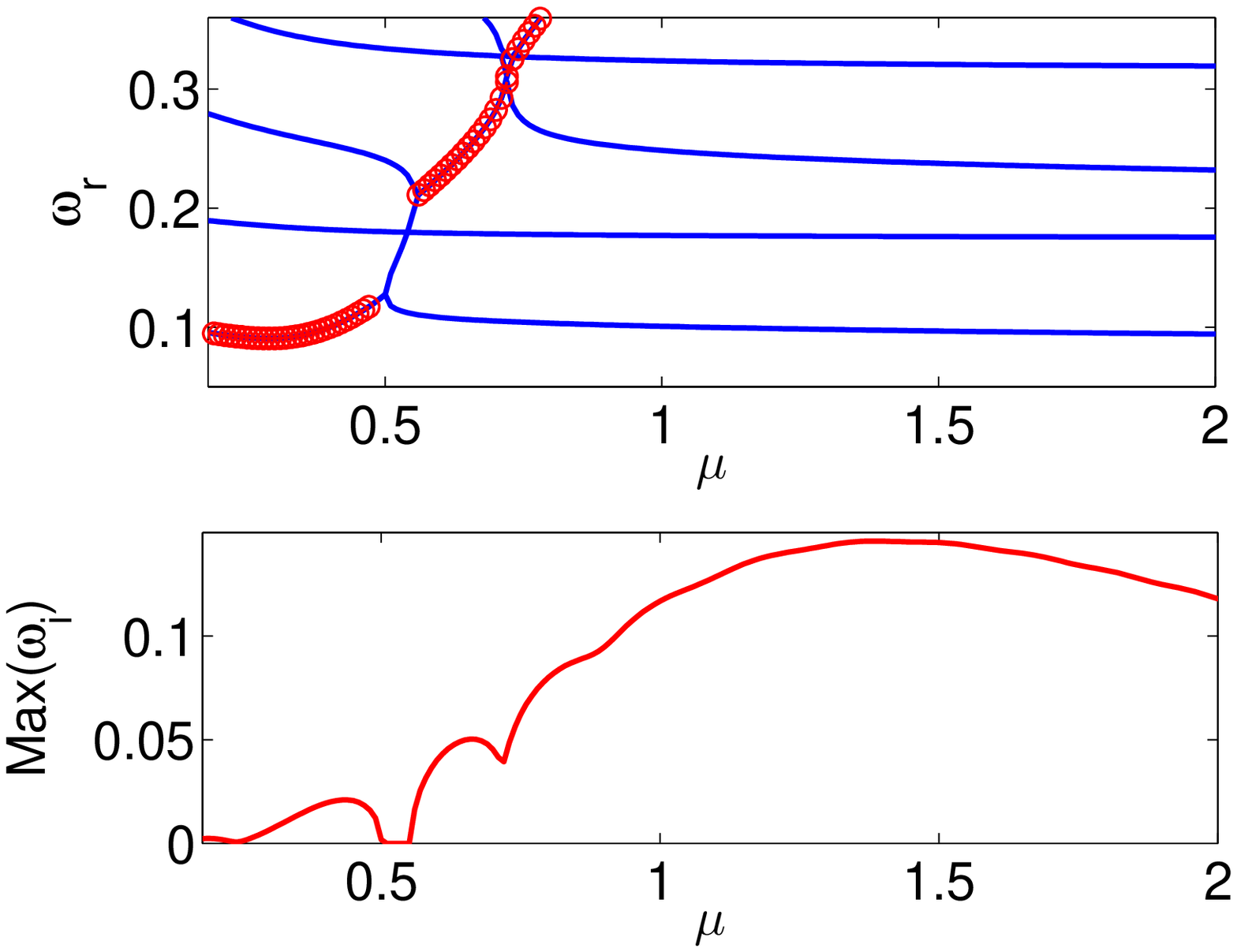}
\includegraphics[scale=0.32]{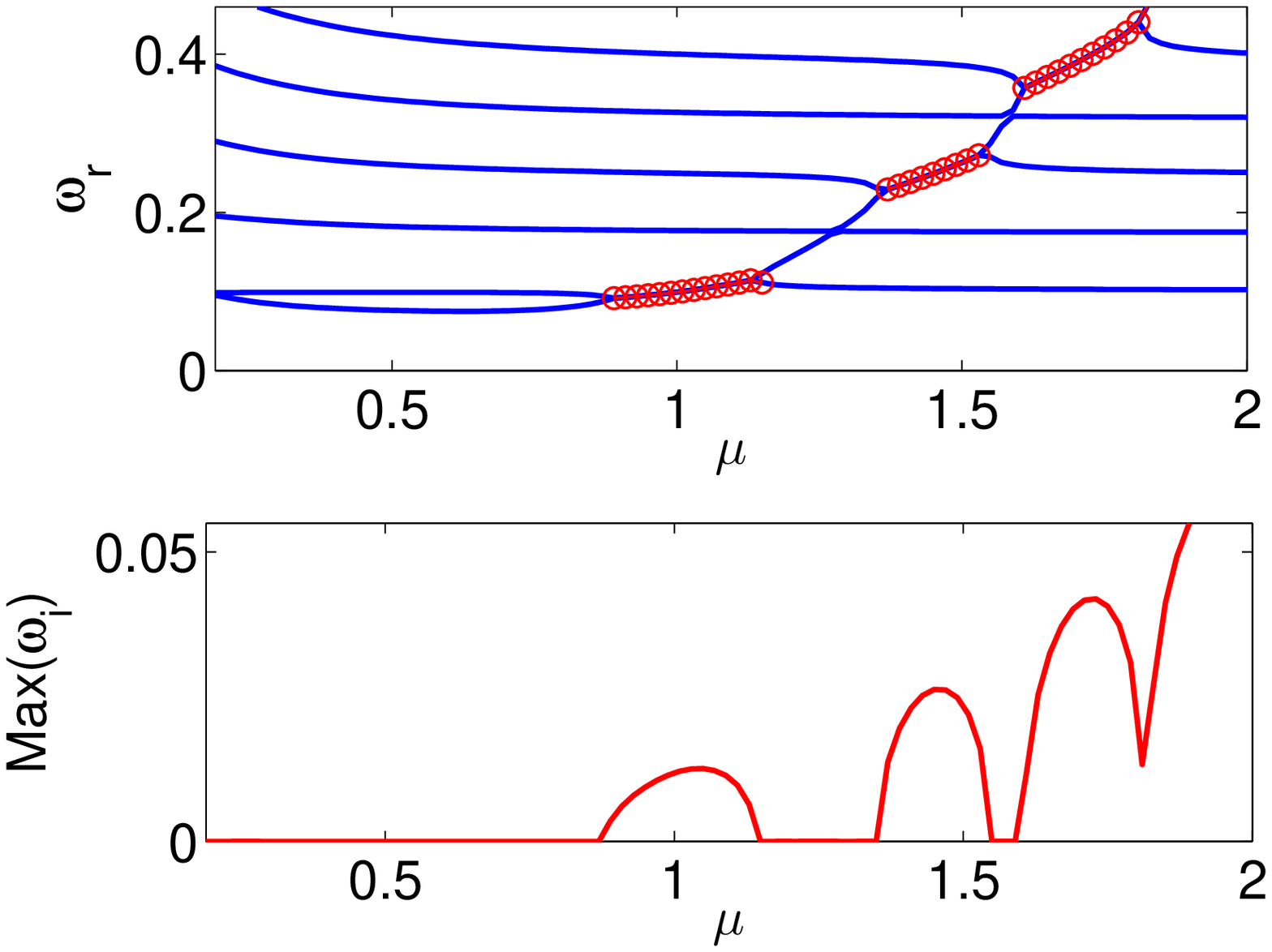}
\includegraphics[scale=0.32]{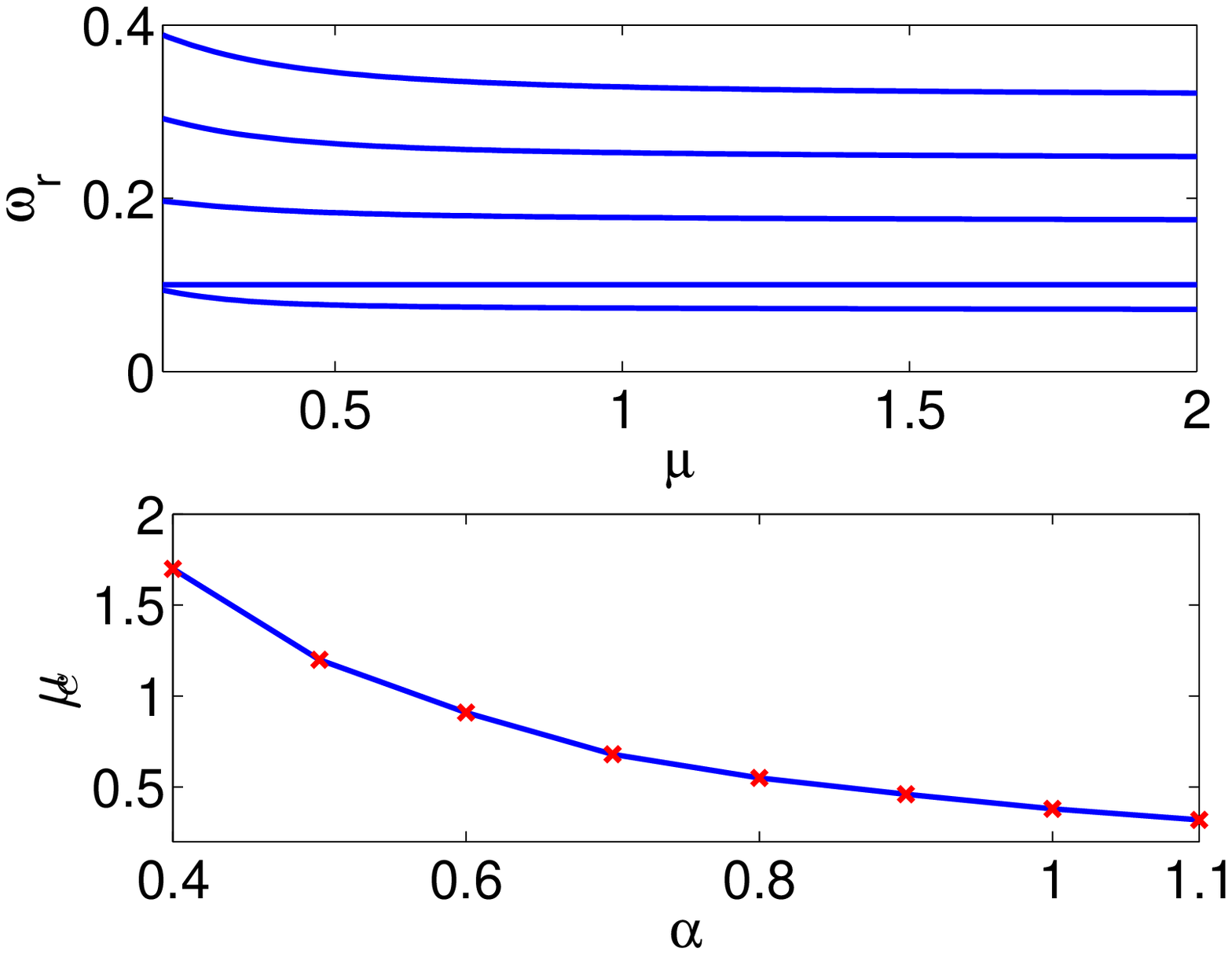}
\caption{(Color online) The linear stability analysis for the first-excited state, corresponding to a single discrete dark soliton. The three top panels show the real part of the lowest-order eigenvalues and the two left bottom panels show the maximum of the imaginary part of the eigenvalues, both as functions of the chemical potential $\mu$. The bottom right panel shows the dependence of the critical value of the chemical potential, $\mu_c$, for the onset of the instability as a function of $\alpha$. Branches shown with circles (in red) in the two top left panels denote dynamically unstable modes, which have emerged upon collision of modes with opposite Krein sign. The parameter values are $\alpha=1.2$ (left column), $\alpha=0.6$ (middle column), and $\alpha=0.1$ (right column); the trap strength is in all cases $\Omega=0.1$.}
\label{e1}
\end{figure*}

Another branch, of slope $\eta=3$, but with the three excited sites featuring an asymmetric configuration (see state C in the third middle panel of Fig.~\ref{energyvsatoms}), also exists for values of $N/\alpha$ smaller than the ones pertaining to (blue) branch B. The states of this branch, which is depicted by a dashed purple curve, are unstable with the corresponding excitation spectra being characterized by a pair of imaginary eigenvalues. The branch C coexists with another one, with two excited sites, namely branch E (depicted by a solid purple curve), which has a slope $\eta=2$. The states belonging to branch E (see, e.g., the example in the bottom left panel of Fig.~\ref{energyvsatoms}) are stable. Both branches C and E continue (as $N$ and $\mu$ are decreased) up to a certain value of the chemical potential, $\mu=0.555$, where they collide and annihilate through a saddle-center bifurcation.
%
%
Notice that still another branch of slope $\eta=2$ exists, namely the
(yellow) branch H, which
starts from the linear limit (at $\mu=0.5$) and remains stable at least up to $\mu=1.2$, as shown in Fig.~\ref{energyvsatoms}.
The states of this branch are anti-symmetric -- see
the example in the bottom right panel of Fig.~\ref{energyvsatoms}.

We now focus on another branch, of slope $\eta=1$ (with the sites $j=0$ excited), which exists
for values of $N/\alpha$ greater than the ones pertaining to the (blue) branch A; an example of a state belonging to this (solid green) branch is state G -- see third bottom panel of Fig.~\ref{energyvsatoms}. As $\mu$ is decreased, the states belonging to this branch are stable down to the value of chemical potential $\mu=0.557$: at this point, the slope $\eta$ changes sign, i.e., $\eta<0$. The change of slope manifests instability according to the slope criterion,  as suggested by the general stability criteria summarized in Section III of Ref. \cite{Wein2008}. It should be noted that instability occurs when either the slope criterion (well known also as a Vakhitov-Kolokolov (VK) criterion \cite{VK}), or the spectral condition \cite{Grillakis1, CKRT1} fails. It is interesting to remark that the Sturm-Liouville-type analysis discussed in Sections III and IV of the present paper, implies that the abstract set-up\cite{Grillakis1, CKRT1} for the implementation of the spectral conditions discussed therein, is valid for our problem. Another interesting observation is that the states of branch G are positive. The positivity suggests the validity of the abstract stability criteria \cite{GSS} for ``positive solitons''  which are applicable in NLS-type systems with linear and nonlinear spatially dependent potentials, and are associated with VK-type slope criteria.

Note that the excitation spectra of the states belonging to the continuation of branch G for $\mu<0.557$ are characterized by a pair of imaginary eigenfrequencies.

Next, a decrease of $\mu$ (and increase of $N/\alpha$) up to the point $\mu=0.55$ results in a pitchfork bifurcation, although this is less transparent in the variables illustrated in the bifurcation diagram of the top right of Fig.~\ref{energyvsatoms} (see also below).
The three branches resulting from this symmetry-breaking bifurcation are the asymmetric branch F (dashed green line), its parity-symmetric one -- which
has the same atom number $N/\alpha$ -- and branch D (dotted green line). The symmetry-broken branch F inherits the stability of the original
branch from which it stemmed (i.e., branch G), while the symmetric
continuation of branch G, namely branch D is, in fact, further destabilized, with the excitation spectra of the pertinent states being characterized by two pairs of imaginary eigenfrequencies.
The above mentioned pitchfork bifurcation is clearly illustrated in Fig.~\ref{beef}, where the difference $\psi_1-\psi_{-1}$ of the two outer sites is plotted as a function of the normalized atom number $N/\alpha$ (the notation,  in terms of the use of dashed and dotted lines, is the same to the one used in the top right panel of Fig.~\ref{energyvsatoms}).

It should be remarked that in general the modes discussed in Fig.~\ref{energyvsatoms} cannot be expressed analytically. On the one hand, analytical expressions could be derived under the assumption that these structures are, in fact, compactly supported, i.e., that the solution is only supported on a few (e.g. three) sites. It is not hard to see that this is not true, by considering the
equation of the "first vanishing" site. Hence, such an assumption is not self-consistent.
Even if we bypass the above nontrivial concern and we assume three nontrivial sites and
symmetry, we may inherit two cubic equations for the stationary solution
elements  which will result ultimately in a $6$th order algebraic equation.
Even if such an equation is solvable, the analytical expressions involved are so tortuous that there is no significant intuition to be gained from this process.

Concluding this section, it is important to notice that the study of the rich bifurcation structure presented above highlights the existence of nonlinear states (such as the ones corresponding to the branches C, F and E in Fig.~\ref{energyvsatoms}) without a linear counterpart.

\section{Stability of the nonlinear states}
\subsection{The first-excited state}

As previously discussed, the
ground-state of the system has been found in the discrete case to be always
stable (i.e., for every value of $\alpha$). On the other hand, as concerns
the stability of the excited nonlinear states
(pertaining to discrete dark multi-solitons as the nonlinearity increases),
we note the following. Since we are interested in investigating the effect of discreteness on the stability of these states, we have performed the BdG analysis for three different values of the lattice spacing $\alpha$ (and fixed value of the trap strength, $\Omega=0.1$). In particular, we have considered the following cases: $\alpha=1.2$ (corresponding to a strongly discrete case), $\alpha=0.6$ (corresponding to a moderate discreteness), and $\alpha=0.1$ (corresponding to a nearly-continuum setting); respective results are shown in Figs.~\ref{e1} and \ref{e2} for the first- and second-excited state, respectively.

In the top left panel of Fig.~\ref{e1} we show the real part of the lowest (four) eigenvalues, while in the bottom left panel we show the maximum of the imaginary part of the eigenvalues, for $\alpha=1.2$ (strongly discrete case). The branch indicated with circles (in red) has emerged from the collision of two modes with opposite Krein signs, namely the anomalous mode and the lowest positive energy mode, and it is unstable up to the value $\mu=0.5$ of the chemical potential. This unstable branch collides with the next mode producing no instability, but from the value  $\mu=0.55$ onwards the anomalous mode starts colliding with higher-order modes, thus producing a new branch which is unstable for all values of $\mu>0.55$. Accordingly, a small stability window is shown to form in the bottom left panel of Fig.~\ref{e1} (for $0.5<\mu<0.55$). For a smaller value of the lattice spacing, $\alpha=0.6$ (i.e., for moderate discreteness), the collision between the anomalous and the first positive energy mode occurs for a larger value of chemical potential, i.e., for $\mu=0.86$; thus, the configuration is initially stable, then it is characterized by an instability window for $0.86 <\mu<1.15$, and it becomes again stable for $1.15 <\mu<1.35$ (see bottom middle panel of Fig.~\ref{e1}). After a small stability window (for $1.55 <\mu<1.6$), the anomalous mode continuously collides with higher-order modes resulting to instability for all values of $\mu>1.6$. Notice that the existence of such (in)stability windows was reported in Ref.~\cite{g1}, where a similar BdG analysis was performed (solely for the single discrete dark soliton configuration in a harmonic trap).

It is important to note that for an even smaller value of the lattice spacing, i.e., for $\alpha = 0.1$ (close to a nearly continuum configuration), the anomalous mode never collides with the first positive energy mode and, thus, the first-excited state (corresponding to a quasi-continuum single dark soliton) is always stable -- see top right panel of Fig.~\ref{e1} -- in accordance with the findings of Refs.~\cite{mur,carr}. The same result can also be concluded by the bottom right panel of Fig.~\ref{e1}, where the critical value $\mu_c$ of the chemical potential for the onset of instability (i.e., for the collision between the anomalous and Kohn modes) is a monotonically decreasing function of the lattice spacing $\alpha$: this indicates that (instability) stability is expected in the (discrete) continuous limit of the model.

\begin{figure}[tbp]
\includegraphics[scale=0.4]{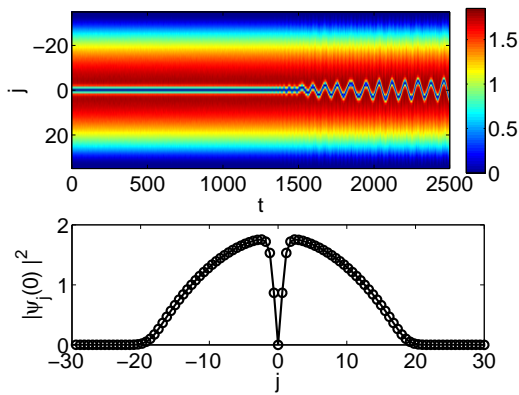}
\caption{(Color online) The top panel shows a spatio-temporal contour plot of the density of the first-excited state (corresponding to a single discrete dark soliton), for parameter values $\mu=1$, $\alpha=1.2$, and $\Omega=0.1$. The soliton stays at rest, up to to $t\approx 1500$, and then starts to perform oscillations of growing amplitude. The bottom panel shows the initial density profile.}
\label{s1}
\end{figure}

We conclude the study of the stability of the first excited state by investigating the dynamics of unstable configurations. In particular, in Fig.~\ref{s1}, we show the evolution of an unstable discrete dark soliton, corresponding to parameter values $\mu=1$, $\alpha=1.2$, and $\Omega=0.1$, as obtained by direct numerical integration of the DNLS Eq.~(\ref{eq1}). The initial condition, chosen in an unstable region with a relatively high instability growth rate (see bottom left panel of Fig.~\ref{e1}), is a discrete dark soliton shown in the bottom panel of Fig.~\ref{s1}. As shown in the top panel of Fig.~\ref{s1}, the discrete dark soliton is at rest up to $t\approx 1500$; then, the instability sets in (due to the numerically-induced noise generation) and the soliton starts to perform oscillations of growing amplitude -- a typical scenario occurring when a dark soliton is subject to an oscillatory instability (see, e.g., Ref.~\cite{g1}).

\begin{figure*}[tbp]
\includegraphics[scale=0.4]{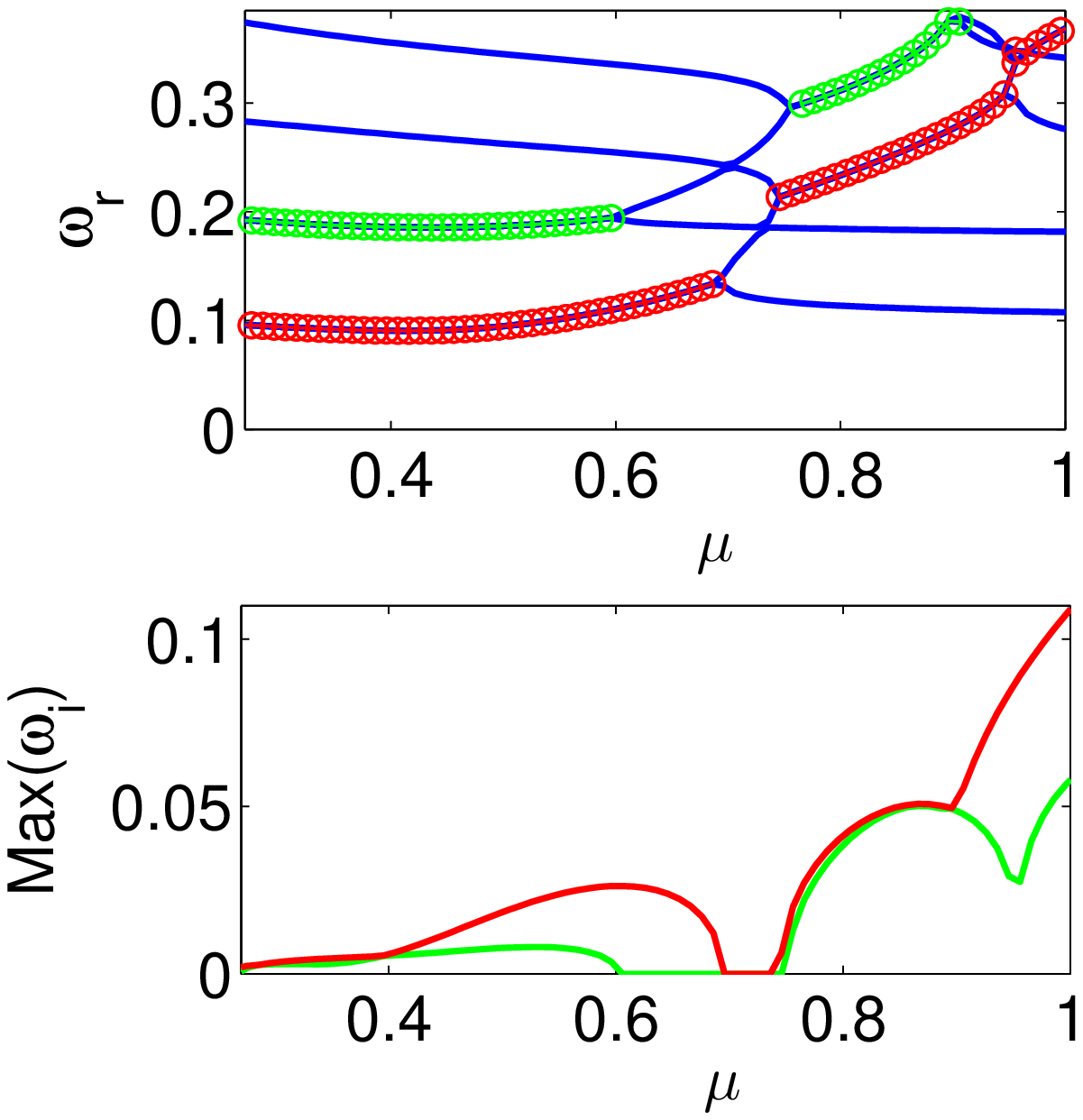}
\includegraphics[scale=0.4]{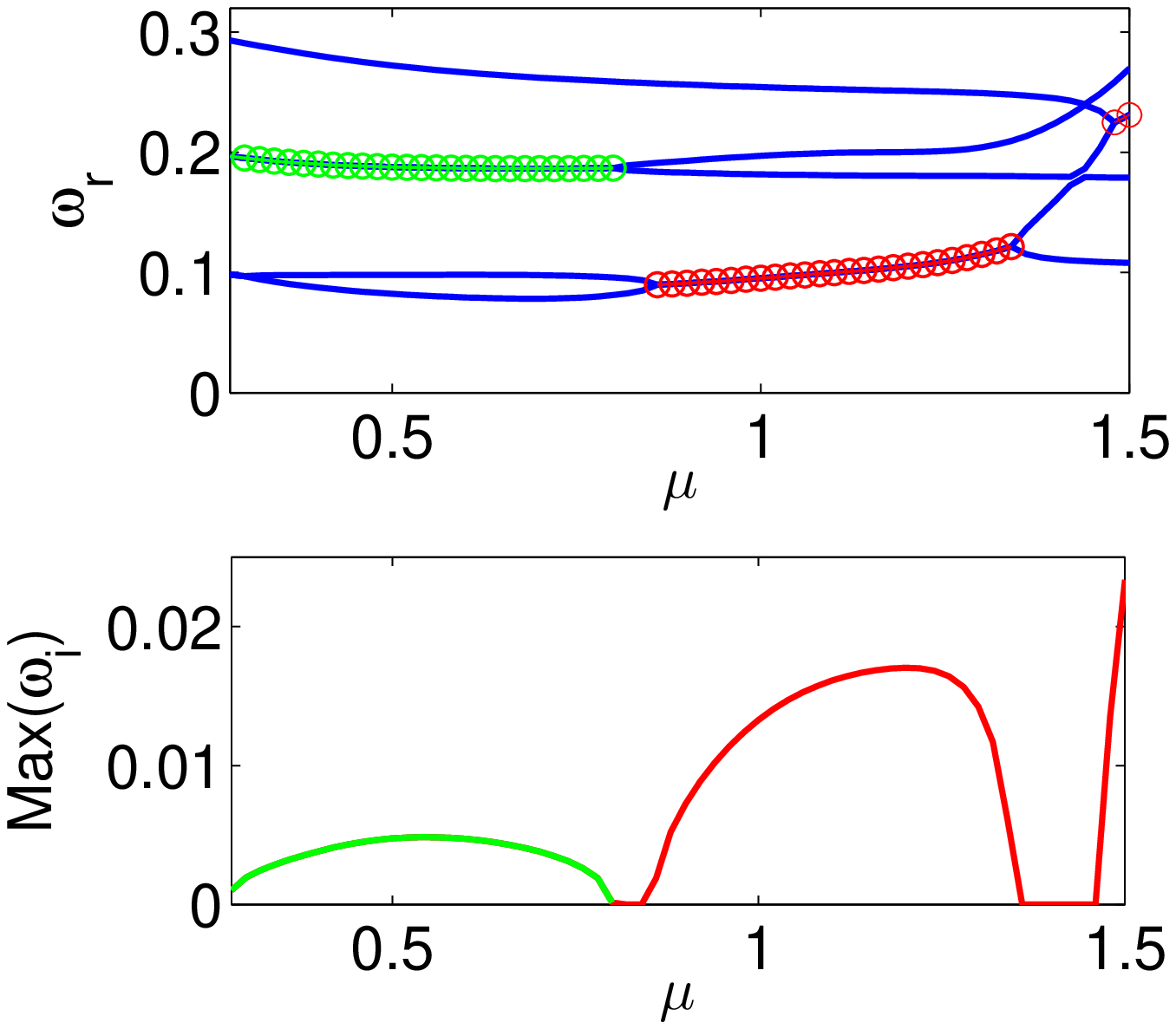}
\includegraphics[scale=0.4]{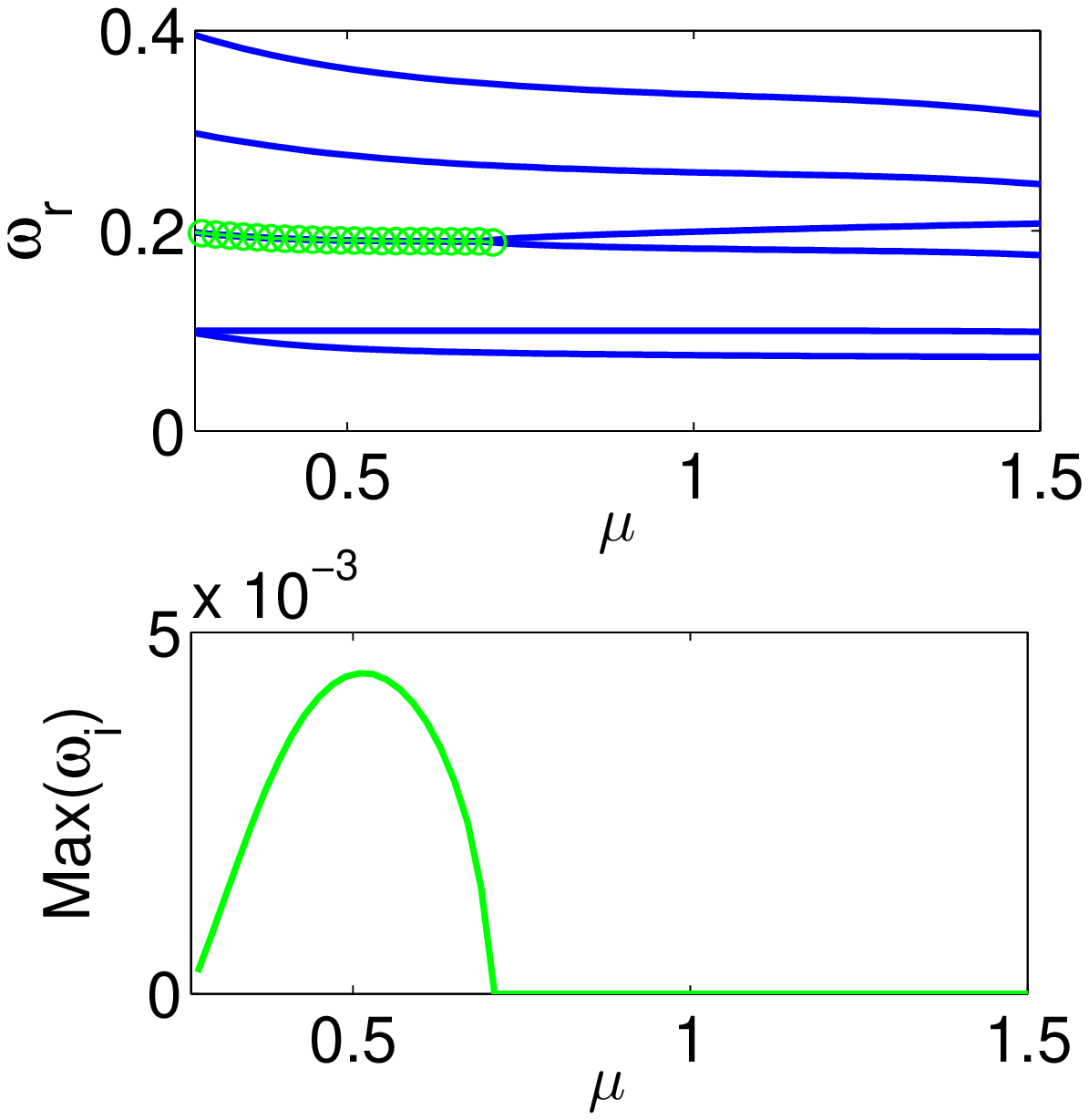}
\caption{(Color online) The BdG analysis for the second-excited state, corresponding to a discrete dark soliton pair.
The top (bottom) panels show the real (imaginary) part of the lowest-order eigenvalues as functions of the chemical potential $\mu$. Branches shown with circles (in red or green) in the top panels denote dynamically unstable modes, which have emerged upon collision of modes with opposite Krein sign. The parameter values are $\alpha=1.2$ (left column), $\alpha=0.6$ (middle column), and $\alpha=0.1$ (right column); the trap strength is in all cases $\Omega=0.1$.}
\label{e2}
\end{figure*}

\subsection{The second-excited state}

We proceed with the stability analysis of the second-excited state, corresponding to a dark soliton pair; our basic results are presented in Fig.~\ref{e2}. First, we note that a fundamental difference of this case with the previous one is the existence of a second anomalous mode (recall that the number of anomalous modes in the excitation spectrum equals the number of dark solitons \cite{multi,zezyulin,Law}). The first anomalous mode (the one with the smaller eigenfrequency corresponding to the {\it in-phase motion} of the two dark solitons \cite{multi} -- see the red branches in the top panels of Fig.~\ref{e2}) follows a behavior
similar to the one found in the single-dark soliton state. Thus, the discreteness induced instability presented in the previous case persists also in the two-soliton configuration. As concerns the behavior of the second anomalous mode (the one with the larger eigenfrequency corresponding to the {\it out-of-phase motion} of the two dark solitons \cite{multi}-- see the green branches in the top panels of Fig.~\ref{e2}) we note the following. Starting with the left column panel (for $\alpha=1.2$), it is observed that the second anomalous mode initially resonates with the second positive energy mode, and an unstable quartet of eigenfrequencies (depicted in green) emerges. This quartet persists up to $\mu=0.6$, where the two modes split. This way, a stability window is formed (see the bottom left panel of Fig.~\ref{e2}) which, however, is effectively reduced by the instability induced from the first anomalous mode; in fact the stability window corresponds to $0.7<\mu<0.74$, an interval defined by the unstable branch corresponding to the first anomalous mode (compare the red and green lines in the bottom left panel of Fig.~\ref{e2}). Next, the second anomalous mode collides with a higher-order mode producing no instability but, eventually, further collisions with higher modes lead to instability.

For a smaller value of the lattice spacing ($\alpha=0.6$), and contrary to the previously examined -- highly discrete -- case of $\alpha=1.2$, the first anomalous mode is initially stable, but becomes unstable for $\mu=0.85$. On the other hand, the quartet that has emerged from the second anomalous mode and the second  positive energy mode (which is initially unstable as before) splits at $\mu=0.8$; this way, a small stability window is created for $0.8<\mu<0.85$ (see bottom middle panel of Fig.~\ref{e2}), while for $\mu>0.85$ the configuration is dynamically unstable. Finally, for $\alpha=0.1$ (corresponding to a quasi-continuum configuration), the right column panels of Fig.~\ref{e2} suggest that an instability induced by the second anomalous mode occurs for $\mu<0.7$, but then, for $\mu>0.7$ the configuration remains stable (although for sufficiently large $\mu$
it will become unstable again). This result is in accordance with the findings of Refs.~\cite{multi,pelink1}, which suggest that in the continuum limit the multi-soliton solution is unstable near the linear limit (due to the second anomalous mode) and may only be unstable thereafter in parametric windows due to collisions of the second anomalous mode with higher positive energy ones.

\begin{figure*}[tbp]
\includegraphics[scale=0.32]{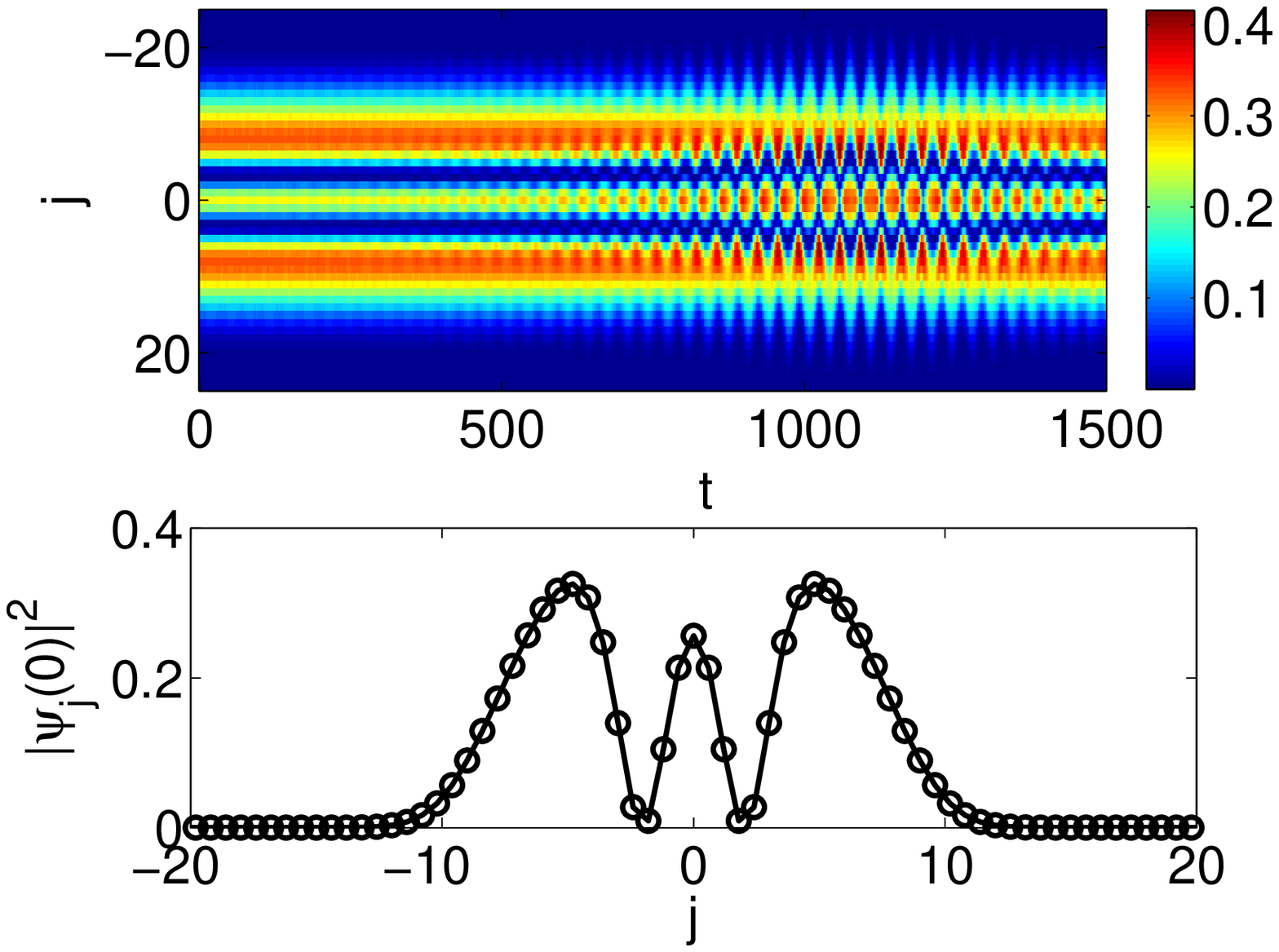}
\includegraphics[scale=0.32]{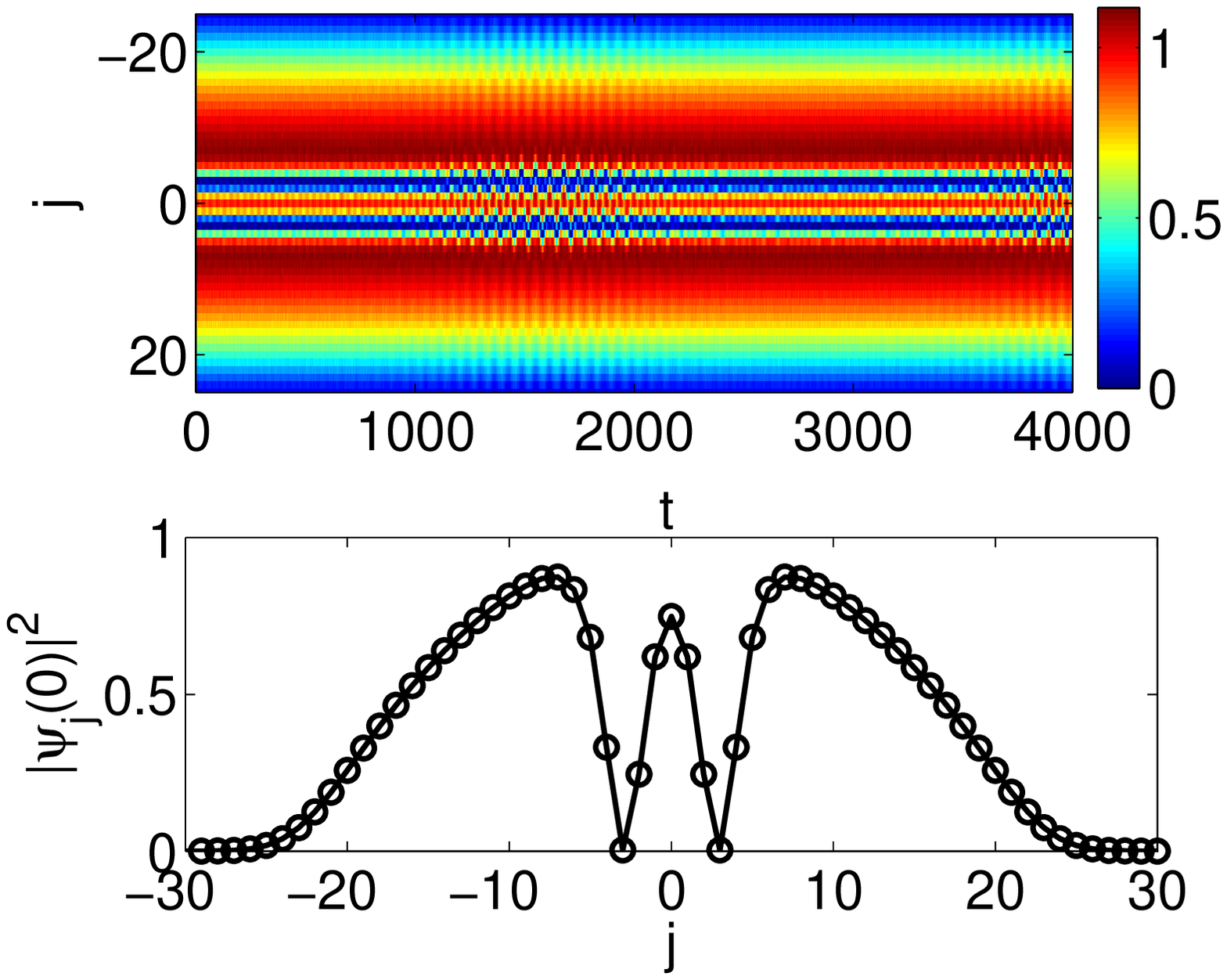}
\includegraphics[scale=0.32]{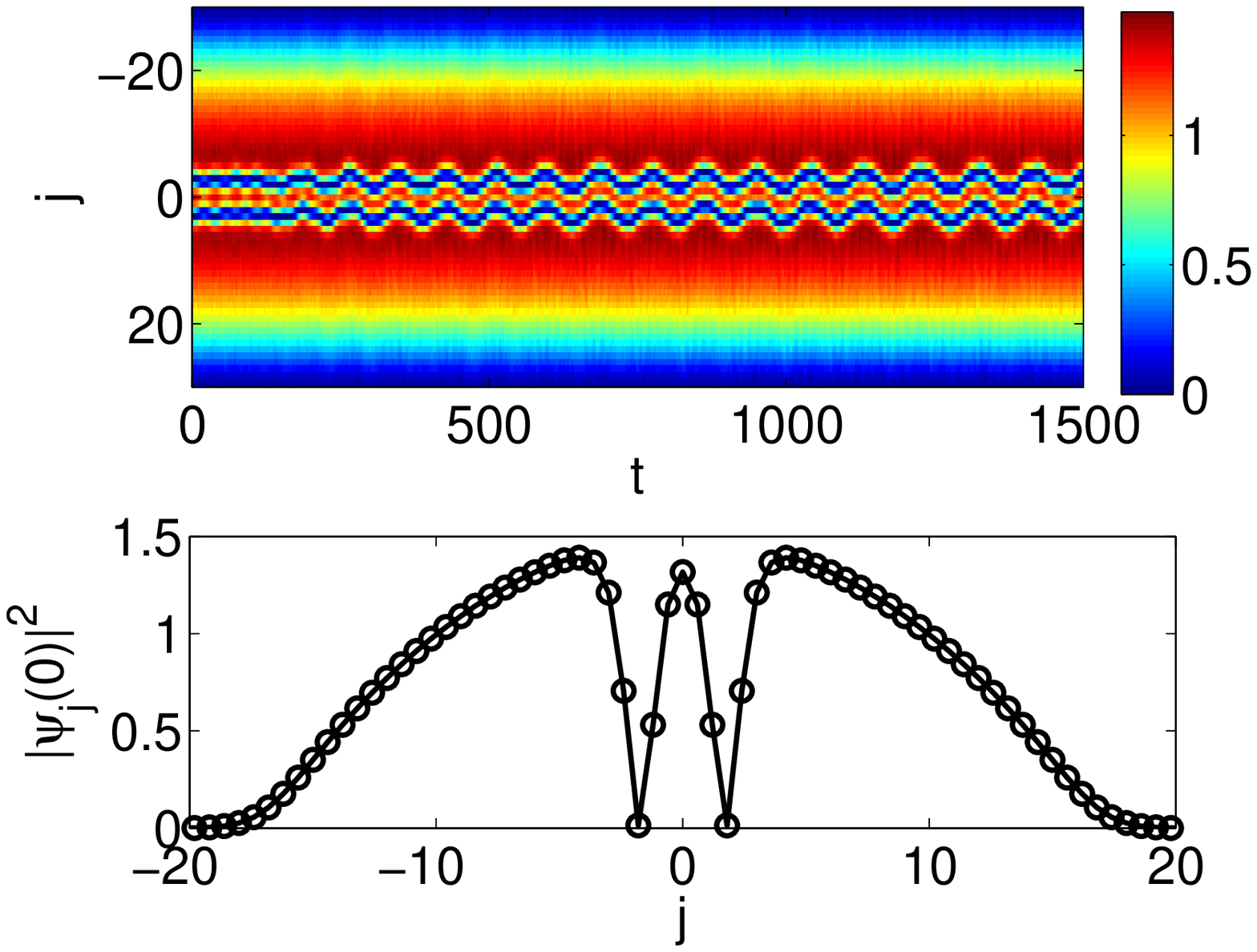}
\caption{(Color online) Same as Fig.~\ref{s1}, but for the second-excited state, corresponding to a discrete dark soliton pair, for parameter values $\mu=0.5$ (corresponding to the first instability band -- see bottom middle panel of Fig.~\ref{e2}), $\alpha=0.6$, and $\Omega=0.1$.}
\label{s2}
\end{figure*}

In the left set of panels of Fig.~\ref{s2} we show the dynamics of an unstable two-dark soliton configuration, corresponding to parameter values $\mu=0.5$, $\alpha=0.6$, and $\Omega=0.1$, as obtained by direct numerical integration of the DNLS of Eq.~(\ref{eq1}). The initial condition is a discrete two-dark soliton state shown in the bottom left panel of Fig.~\ref{s2}. As shown in the top left panel of Fig.~\ref{s2}, the discrete dark soliton pair is stationary up to $t\approx 500$; then, the instability manifests itself: the out-of-phase motion of the two-soliton state excites the second positive energy mode (often referred to as quadrupole mode in the continuum case \cite{book2}) of the system. This excitation results in a breathing behavior of the configuration. This type of instability was also found in the continuum counterpart of the system (see Fig. 5(c) of Ref.~\cite{multi}). Note that this particular simulation corresponds to the first instability band shown in the bottom middle panel of Fig.~\ref{e2}. The results of simulations performed with values of $\mu$ corresponding to the second and third instability bands, i.e., for $\mu=1$ and $\mu=1.5$, are respectively shown in
the middle and right panels of Fig. \ref{s2}. In both cases, the initially quiescent two-dark-soliton configuration becomes unstable and is set into motion, with the solitons performing an in-phase motion. It is clearly observed that the instability manifests itself at different times in the two cases, namely at $t \approx 800$ and $t \approx 180$ for
the middle and right panels, respectively; furthermore, the amplitude of oscillation of the dark soliton configuration in the former case is much smaller than the one shown in the latter. Thus, although both cases correspond to  linearly unstable two-dark-soliton configurations, the one shown in
the middle panel appears to be more robust than the one in the right.
This may be partially connected to the instability growth rates, but
perhaps, more importantly, to the different modes of the background with
which the internal in-phase soliton mode resonates in the different cases.

\section{Conclusions}

In this work, we presented a systematic study of the existence, stability and bifurcations of nonlinear states of a self-interacting quantum harmonic oscillator (QHO) on a lattice. The considered model, namely a discrete NLS equation incorporating a (discrete) harmonic trap, may be used to describe the dynamics of an array of BEC droplets in a deep optical lattice, but also discrete nonlinear guided-wave optical systems.

Our considerations started with the analysis of the pertinent linear problem. We presented the energy spectrum and the eigenstates of the linear problem as functions of the lattice spacing $\alpha$. This way, we spanned all possible cases, starting from the continuum limit (i.e., the well-known QHO for $\alpha\rightarrow 0$) to the anti-continuum one ($\alpha\rightarrow\infty$), where the ground state energy asymptotes to zero, while the excited states exhibit a parabolic energy spectrum. Next, using global bifurcation theory (and employing, in particular, a functional-analytic theorem from the work of Rabinowitz), we rigorously proved that -- in the discrete regime -- all eigenstates of the linear problem can be continued to nonlinear ones, so that each linear state possesses a nonlinear counterpart. Using this result, we were able to construct numerically the nonlinear states emerging from their linear siblings; this way, we found the ground state of the system (which acquires the Thomas-Fermi profile in the anti-continuum limit), while the excited states take the form of a chain of stationary discrete dark solitons.
The anti-continuum limit was studied as well;
it was found
that the solutions present a complex bifurcation
structure, which was elucidated along with the stability of the
corresponding branches. The pertinent bifurcation diagram also revealed the existence of nonlinear states with no linear counterpart.

We also performed a detailed linear stability analysis of the different
nonlinear solutions ensuing for different values of the lattice
parameter $\alpha$, solving the BdG equations eigenvalue problem, for the ground state, as well as for the first and second excited states (the latter, correspond to a single dark soliton and a pair of dark solitons, respectively, in the strongly nonlinear regime). While the ground state was found to be completely stable for all values of the lattice spacing, this was not the case for the discrete dark soliton states, which revealed a quite rich stability spectrum. In the strongly discrete regime, the single dark soliton was found to be potentially unstable, due to collisions of the
first anomalous mode eigenvalue with the rest of the normal modes of the
system, for increasing chemical potential $\mu$. As the system becomes more
continuous, i.e., for decreasing lattice spacing $\alpha$, stable windows
appear and gradually expand; eventually, in the quasi-continuum regime, the anomalous mode remains
below the positive energy mode for all values of $\mu$, and the soliton
becomes stable. A similar behavior (from the strongly discrete to the
quasi-continuum regime) but with additional sources of potential
instabilities (from the additional anomalous mode) was
identified for the two dark soliton configuration.
The anomalous mode responsible for out-of-phase motion between the
solitary waves is initially in resonance
with the second positive energy mode, thus creating an instability, but
eventually they split to create small windows of stability (in the discrete
regime). In the quasi-continuum limit, the first anomalous modes yields
no instabilities while the second one leads to windows of instability
(which are more pronounced near the linear limit).

In both cases, our analysis revealed that the effect of discreteness is to
chiefly offer additional sources of instability of the dark soliton states
for atom number parameter ranges for which they would be
in the continuum counterpart of the model.
This is due to the pronounced dependence of soliton anomalous modes
on the chemical potential, as well as due to the discreteness eliminating
some of the symmetries (such as the dipolar symmetry of the first positive
 energy mode) present in the continuum limit.

We would like to conclude by mentioning some main differences between the results concerning (\ref{eq111}) and (\ref{eq3}), and the DNLS equation without potential\cite{Kon07}. One of these differences concerns the infinite lattice limit, especially in terms of the bifurcation analysis carried out herein. The case of the parabolic potential retains its point-spectrum nature even in that limit and the spacing of the energy
levels is chiefly controlled by the trap frequency $\Omega$. On the contrary, in the absence of $\Omega$\cite{Kon07} as the lattice becomes
infinite in the realm of the above paper, the point spectrum due to the finiteness of the domain
converts itself into a continuous spectral band and hence its properties (and bifurcations) are
substantially different. The bifurcation mechanism analyzed herein (bifurcation from simple eigenvalues) is one of the main types for generation of nonlinear states, as it has been highlighted in Ref. \cite{Wein2008} (see Section II\cite{Wein2008}, pg. 046602-2).
Another relevant difference concerns the lengthscales of these states. In the absence of $\Omega$,
the point spectrum eigenfunctions are spatially "extended" (within the length-scale of the lattice).
On the other hand, the spatial eigenfunctions in the case of the parabolic trap problem are localized within
a lengthscale controlled by the trap frequency $\Omega$. Hence, the presence of a parabolic trap
yields an additional lengthscale which can be used to induce interesting phenomena, such as for example
the ones that emerge from the competition of the trap lengthscale with the intrinsic lengthscale of
the lattice. This is e.g. what produces the complex bifurcation diagrams such as the one of Fig.~\ref{energyvsatoms},
while such a phenomenology is likely more limited in the context of a $2$- or a $3$-site lattice (only).

It would be interesting to extend our considerations in other settings,
such as ones involving different types of trapping potentials,
multi-dimensional one-component systems (e.g., in the case of
both dark soliton and vortex type entities in two-dimensional
settings), as well as in multi-component systems. Work is in progress in
these directions and
relevant results will be presented in future publications.

\section*{Acknowledgments} G.T. acknowledges support from the Alexander S. Onassis Foundation.
P.G.K. gratefully acknowledges support from NSF-DMS-0349023, NSF-DMS-0806762, NSF-CMMI-1000337, and from
Alexander von Humboldt and
Alexander S. Onassis Foundations.
The work of F.K.D. and D.J.F. was partially supported by the Special Account for Research Grants of the
University of Athens.


\section*{Appendix}
\label{App}
In this section we give a proof on the existence of an excitation threshold in the sense of Ref.~\cite{Wein99}, for the DNLS equation (\ref{eq1}) considered in the lattice $\mathbb{Z}^\mathcal{N}$, $\mathcal{N}\geq 1$. For technical purposes it is more convenient to work with the focusing version of DNLS of Eq. (\ref{eq1}), having {\em the opposite sign on the nonlinearity}. We shall reduce the DNLS of (\ref{eq1}) to the one with an effectively opposite coefficient
of the nonlinearity under the, so-called,  {\em staggering transformation}. This transformation is defined as (see, e.g., Ref.~\cite{Panos3}),
\begin{eqnarray}
\label{stag}
\psi_j\rightarrow (-1)^{p}\psi_j,\;\;\; p=\sum_{i=1}^\mathcal{N} j_i,
\end{eqnarray}
for
$j:=(j_1,j_2,\ldots,j_\mathcal{N})\in\mathbb{Z}^\mathcal{N}$ (a trivial multiplication
by a suitable phase factor is also needed to form the corresponding local term
within the discrete Laplacian). Thus, under Eq.~(\ref{stag}), the $\mathcal{N}$-dimensional, focusing version of the system (\ref{eq1}) with a general power nonlinearity, $|\psi_j|^{2\sigma}\psi_j$, can be actually written as (taking advantage of the time-reversal symmetry)
\begin{eqnarray}
\label{eqWTf}
i \dot{\psi}_j+\frac{1}{2\alpha^2}\Delta_2 \psi_j + \frac{1}{2} \alpha^2\Omega^2 |j|^2 \psi_j +|\psi_j|^{2\sigma}\psi_j=0,
\end{eqnarray}
considered in the $\mathcal{N}$-dimensional cube of $\mathbb{Z}^{\mathcal{N}}$ with edges of length $2L$,
\begin{eqnarray*}
&&\overline{\mathcal{Q}}=\{(x_{j_1},\ldots,x_{j_{\mathcal{N}}})\,:\,0\leq j_1,\ldots,j_\mathcal{N}\leq K+1\},\\
&&x_{j_i}=-L+j_{i}\alpha,\;\;\alpha=\frac{2L}{K+1},\;\;i=1,\ldots,\mathcal{N}.
\end{eqnarray*}
The discrete eigenfunctions on $\overline{\mathcal{Q}}$ are denoted by $$\psi_j(t)=\psi(x_{j_1},x_{j_2},\ldots,x_{j_{\mathcal{N}}},t).$$
The interior of the cube $\overline{\mathcal{Q}}$ is given by
\begin{eqnarray*}
\mathcal{Q}=\{(x_{j_1},\ldots,x_{j_{\mathcal{N}}})\,:\,1\leq j_1,\ldots,j_\mathcal{N}\leq K\},
\end{eqnarray*}
and (\ref{eqWTf}) is supplemented with Dirichlet boundary conditions
\begin{eqnarray}
\label{eqWT2f}
\psi_j=0,\;\mbox{on}\;\partial\mathcal{Q}:=\overline{\mathcal{Q}}\setminus\mathcal{Q}.
\end{eqnarray}
Solutions $\psi_j \rightarrow \psi_j\exp( -i \mu t)$, of (\ref{eqWTf})-(\ref{eqWT2f}), are equivalently, solutions of the constrained minimization problem
\begin{eqnarray}
\label{eqWT3}
\mathcal{I}_{\mathcal{R}}=\left\{H[\psi]\,:\,N[\psi]=\mathcal{R}\right\},
\end{eqnarray}
where the chemical potential $\mu$ appears as a Lagrange multiplier associated to the minimizer of (\ref{eqWT3}).
In (\ref{eqWT3}), $H$ denotes the Hamiltonian
\begin{eqnarray*}
H[\psi]=\frac{1}{2\alpha^2}(-\Delta_2\psi,\psi)_2&+&\frac{1}{2}\alpha^2\Omega^2\sum_{\overline{\mathcal{Q}}}|j|^2|\psi_j|^2\\
&-&\frac{1}{\sigma+1}\sum_{\overline{\mathcal{Q}}}|\psi_j|^{2\sigma+2},
\end{eqnarray*}
while $$N[\psi]:=\sum_{\overline{\mathcal{Q}}}|\psi_j|^{2},$$ the
norm of the Hilbert space $\ell^2$ of square summable sequences, represents the atom number or optical power (see (\ref{Number})).
We have the following
\begin{proposition}
\label{propWT}
$\mathcal{I}_{\mathcal{R}}\geq 0$ if and only if $\mathcal{R}$ satisfies the inequality
\begin{eqnarray}
\label{eqWT4}
&&\sum_{\overline{\mathcal{Q}}}
|\psi_j|^{2\sigma+2}\leq (\sigma +1)\mathcal{R}^{-\sigma}\left(\sum_{\overline{\mathcal{Q}}}|\psi_j|^2\right)^{\sigma}\nonumber\\
&&\times\left[\frac{1}{2\alpha^2}(-\Delta_2\psi,\psi)_2+\frac{1}{2}\alpha^2\Omega^2\sum_{\overline{\mathcal{Q}}}|j|^2|\psi_j|^2\right],
\end{eqnarray}
for all $\psi\in\mathbb{R}^{\mathcal{N}(K+2)}$.
\end{proposition}
\textbf{Proof:} By the definition of $\mathcal{I}_{\mathcal{R}}$ in (\ref{eqWT3}) it follows that $\mathcal{I}_{\mathcal{R}}\geq 0$ if and only if
\begin{eqnarray}
\label{eqWT5}
\frac{1}{\sigma+1}\sum_{\overline{\mathcal{Q}}}
|\psi_j|^{2\sigma+2}&\leq& \frac{1}{2\alpha^2}(-\Delta_2\psi,\psi)_2\nonumber\\
&+&\frac{1}{2}\alpha^2\Omega^2\sum_{\overline{\mathcal{Q}}}|j|^2|\psi_j|^2,
\end{eqnarray}
for all $\psi\in\mathbb{R}^{\mathcal{N}(K+2)}$. Let now $\psi\in\mathbb{R}^{\mathcal{N}(K+2)}$, $\psi\neq 0$ arbitrary, and consider the element
$z=\sqrt{\mathcal{R}}||\psi||_2^{-1}\psi$. Observing that $N[z]=||z||_2^2=R$, by substitution of $z$ in (\ref{eqWT5}) we derive
(\ref{eqWT4}).\ \ $\diamond$

From the Proposition~4.2, p.~680 of Ref.~\cite{Wein99}, it clearly follows that if $\sigma\geq\frac{2}{\mathcal{N}}$, there exist a constant $C>0$ such that for any $\epsilon>0$, the inequality
\begin{eqnarray}
\label{eqWT6}
\sum_{j\in\mathbb{Z}^\mathcal{N}}
|\psi_j|^{2\sigma+2}&\leq& C\left(\sum_{j\in\mathbb{Z}^\mathcal{N}}|\psi_j|^2\right)^{\sigma}\nonumber\\
&&\times\frac{1}{2\alpha^2}(-\Delta_2\psi,\psi)_2,
\end{eqnarray}
holds  for all $\psi\in\ell^2$. Thus, it is an immediate consequence that there exist $C>0$, such that
\begin{eqnarray}
\label{eqWT7}
&&\sum_{j\in\mathbb{Z}^\mathcal{N}}
|\psi_j|^{2\sigma+2}\leq C\left(\sum_{j\in\mathbb{Z}^\mathcal{N}}|\psi_j|^2\right)^{\sigma}\nonumber\\
&&\times\left[\frac{1}{2\alpha^2}(-\Delta_2\psi,\psi)_2
+\frac{1}{2}\alpha^2\Omega^2\sum_{j\in\mathbb{Z}^\mathcal{N}}|j|^2|\psi_j|^2\right],
\end{eqnarray}
for all $\psi\in\ell^2$ and $\sigma\geq\frac{2}{\mathcal{N}}$. We define for brevity the functional
\begin{eqnarray*}
E[\psi]:=\frac{1}{2\alpha^2}(-\Delta_2\psi,\psi)_2+\frac{1}{2}\alpha^2\Omega^2\sum_{j\in\mathbb{Z}^\mathcal{N}}|j|^2|\psi_j|^2.
\end{eqnarray*}
In analogy with Eq.~(4.2), p.~680 of Ref.~\cite{Wein99}, if $C_*$ is the infimum over all the constants for which (\ref{eqWT7}) holds, then $C_*$ it can be characterized as
\begin{eqnarray}
\label{eqWT9}
\frac{1}{C_*}=\inf\frac{\left(\sum_{j\in\mathbb{Z}^\mathcal{N}}| \psi_j|^2\right)^{\sigma}E[\psi]}{\sum_{j\in\mathbb{Z}^\mathcal{N}}
|\psi_j|^{2\sigma+2}}.
\end{eqnarray}
Therefore, the  {\em excitation threshold $\mathcal{R}_{\mathrm{thresh}}$ for the DNLS equation (\ref{eqWTf})} will be defined by a comparison of (\ref{eqWT4}) and (\ref{eqWT7}) as
\begin{eqnarray}
\label{eqWT8}
(\sigma+1)(\mathcal{R}_{\mathrm{thresh}})^{-\sigma}=C_*.
\end{eqnarray}
We conclude with the following theorem.
\begin{theorem}
\label{thresh}
Let $\sigma\geq\frac{2}{\mathcal{N}}$.\\
$\mathrm{A}$. Assume that $||\psi||^2_{2}=\mathcal{R}$. Then
\begin{eqnarray}
\label{siteWein7}
\mathcal{H}[\psi]\geq E[\psi]\left[1-\left(\frac{\mathcal{R}}{\mathcal{R}_{\mathrm{thresh}}}\right)^{\sigma}\right].
\end{eqnarray}
$\mathrm{B}$. If $\mathcal{R}<\mathcal{R}_{\mathrm{thresh}}$ then $\mathcal{I}_{\mathcal{R}}=0$ and there is no ground state minimizer of (\ref{eqWT3}).\\
$\mathrm{C}$. If $\mathcal{R}>\mathcal{R}_{\mathrm{thresh}}$ then $\mathcal{I}_{\mathcal{R}}<0$ and there exists a minimizer of the variational problem (\ref{eqWT3}).
\end{theorem}
{\bf Proof:} $\mathrm{A}$. Let us note first that (\ref{eqWT6}) holds for any $\psi\in\mathbb{R}^{\mathcal{N}(K+2)}$ with the same optimal constant $C_*$, since  $\mathbb{R}^{\mathcal{N}(K+2)}$ is a finite dimensional subspace of $\ell^2$. Then, using (\ref{eqWT6}) with its best constant $C_*$ we derive that
\begin{eqnarray*}
H[\psi]&=&E[\psi]
-\frac{1}{\sigma+1}\sum_{\overline{\mathcal{Q}}}|\psi_j|^{2\sigma+2}\\
&\geq&E[\psi]-(\mathcal{R}_{\mathrm{thresh}})^{-\sigma}\mathcal{R}^{\sigma}E[\psi],
\end{eqnarray*}
thus (\ref{siteWein7}).\newline
$\mathrm{B}$. Assuming that $\mathcal{R}<\mathcal{R}_{\mathrm{thresh}}$, it follows from (\ref{siteWein7}) that $\mathcal{I}_\mathcal{R}\geq 0$. On the other hand, we may consider some $\tilde{\psi}\in\mathbb{R}^{\mathcal{N}(K+2)} $ such that
\begin{eqnarray*}
||\tilde{\psi}||_{\ell^2}&=&\frac{\sqrt{\mathcal{R}}}{\lambda},\;\;\mbox{where $\lambda>0$ arbitrary}.
\end{eqnarray*}
Considering the element $z_{\lambda}=\sqrt{\mathcal{R}}||\tilde{\psi}||_{2}^{-1}\tilde{\psi}$ we observe that $$||z_{\lambda}||_{\ell^2}^2=\mathcal{R}$$ and
\begin{eqnarray*}
H[z_{\lambda}]=\lambda^2E[\hat{\psi}]-\frac{\lambda^{2\sigma+2}}{\sigma+1}\sum_{\overline{\mathcal{Q}}}|\tilde{\psi}_j|^{2\sigma+2}.
\end{eqnarray*}
For $\lambda$ sufficiently large,  we get that $H[z_{\lambda}]<0$. Therefore if $\mathcal{R}<\mathcal{R}_{\mathrm{thresh}}$ we should have $\mathcal{I}_\mathcal{R}=0$. Assuming that this infimum is attained at a state $\hat{\phi}$, then $\mathcal{I}_\mathcal{R}=0$ implies that
\begin{eqnarray}
\label{siteWein8}
E[\hat{\phi}]&=&\frac{1}{\sigma+1}\sum_{\overline{\mathcal{Q}}}|\hat{\phi}_n|^{2\sigma+2},\\
N[\hat{\phi}]&=&\sum_{\overline{\mathcal{Q}}}|\hat{\phi}_j|^2=\mathcal{R}.\nonumber
\end{eqnarray}
Then, inequality (\ref{eqWT6}) with its best constant $C_*$, if inserted into (\ref{siteWein8}), is giving the contradiction
\begin{eqnarray*}
E[\hat{\phi}]\leq
\frac{1}{\sigma+1}\sum_{\overline{\mathcal{Q}}}|\hat{\phi}_j|^{2\sigma+2}
\leq\frac{1}{2\alpha^2}\left(\frac{\mathcal{R}}{\mathcal{R}_{\mathrm{thresh}}}\right)^{\sigma}
<E[\hat{\phi}].
\end{eqnarray*}
$\mathrm{C}$. By the definitions (\ref{eqWT9}), (\ref{eqWT8}) of $C^*$ and $\mathcal{R}_{\mathrm{thresh}}$ respectively,  it follows that if  $\mathcal{R}>\mathcal{R}_{\mathrm{thresh}}$ then  a $\phi^*\in\ell^2$ should exist which does not satisfy inequality (\ref{eqWT4}), hence $\mathcal{I}_{\mathcal{R}}<0$. Indeed, such a minimizer exists since $H[\psi]$ is bounded from below and in the finite dimensional space $\mathbb{R}^{\mathcal{N}(K+2)}$ the infimum $\mathcal{I}_{\mathcal{R}}<0$ is attained.\ \ $\diamond$

Let us note that the complementary results presented in this section are of independent interest since they prove existence of nonlinear states for the DNLS Eq.~(\ref{eq1}) directly, together with the existence of a threshold for their activation energy. It is important to note {\em that in the continuous limit $\alpha\rightarrow\infty$, an excitation threshold exists only in the critical case $\sigma=\frac{2}{N}$} (see Sections 3 and 4 of Ref.~\cite{Wein99}). We also remark that the results can be extended in the case of the infinite lattice $\mathbb{Z}^{\mathcal{N}}$ by implementing the concentration compactness arguments \cite{Wein99,JNF2010}.

\end{document}